\documentclass[a4paper,12pt]{article}
\usepackage[utf8]{inputenc}
\usepackage{cancel}
\usepackage{ulem}
\usepackage{amsfonts}
\usepackage{amssymb}
\usepackage{graphicx}
\usepackage{amsmath}
\usepackage{enumerate}
\usepackage{mathtools}
\usepackage{subfig}
\usepackage{color}
\usepackage{tikz}\usetikzlibrary{calc}
\usepackage{float}
\usepackage{here}
\usepackage{cite}
\usepackage{mathrsfs}
\usepackage{float,epsfig}
\usepackage{dcolumn}

\usepackage{graphicx}
\usepackage{bm}
\usepackage{amsmath,amssymb,amsthm}
\usepackage[colorlinks=true,linkcolor=blue,citecolor=red]{hyperref}
\newcommand{\be}{\begin{equation}}
\newcommand{\ee}{\end{equation}}
\newcommand{\bea}{\setlength\arraycolsep{2pt} \begin{eqnarray}}
\newcommand{\eea}{\end{eqnarray}}

\def\0{{\sst{(0)}}}
\def\1{{\sst{(1)}}}
\def\2{{\sst{(2)}}}
\def\3{{\sst{(3)}}}
\def\4{{\sst{(4)}}}
\def\5{{\sst{(5)}}}
\def\6{{\sst{(6)}}}
\def\7{{\sst{(7)}}}
\def\8{{\sst{(8)}}}
\def\sst#1{{\scriptscriptstyle #1}}

\makeatletter \@addtoreset{equation}{section}

\usepackage{multirow}
\definecolor{lime}{HTML}{A6CE39}

\tikzset{>=latex}

\begin{document}
%

\title{\normalsize
{\bf \Large	 Deflection angle and Shadows by   Black Holes in  Starobinsky-Bel-Robinson      Gravity from M-theory}}

\author{ \small   A. Belhaj$^1$\footnote{a-belhaj@um5r.ac.ma},  H. Belmahi$^1$\footnote{hajar\_belmahi@um5.ac.ma},  M. Benali$^1$\footnote{mohamed\_benali4@um5.ac.ma},   Y. Hassouni$^1$\footnote{y.hassouni@um5r.ac.ma}, M. B. Sedra$^2$ \footnote{ Authors in alphabetical order.}
	\hspace*{-8pt} \\
	{\small  $^1$ D\'{e}partement de Physique, Equipe des Sciences de la mati\`ere et du rayonnement, ESMaR}\\
{\small   Facult\'e des Sciences, Universit\'e Mohammed V de Rabat, Rabat,  Morocco} \\	
	{\small   $^2$ Material and subatomic physics laboratory, LPMS,   University of  Ibn Tofail, Kenitra,  Morocco} 	} \maketitle

\begin{abstract}
Motivated by M-theory compactifications, we investigate optical properties of black holes in  the   Starobinsky-Bel-Robinsion  gravity.
  Precisely, we  study   the shadows and the deflection angle of light rays by non-rotating and rotating  black holes  in such a novel gravity.   We start by  discussing   the shadows of the Schwarzschild-type solutions. As expected, we  obtain perfect circular shadows  where the size decreases with   a stringy    gravity  parameter   denoted by   $\beta$. We show that this parameter  is  constrained by  the  shadow existence. Combining  the  Newman-Janis algorithm and  the Hamilton-Jacobi mechanism,  we   examine   the shadow  behaviors of  the rotating solutions in terms of  one-dimensional real curves.  Essentially,    we    find    various   sizes and shapes depending  on  the rotating  parameter and the stringy  gravity parameter $a$  and $\beta$, respectively.    To  inspect  the shadow geometric deformations,   we investigate the astronomical observables and   the energy emission rate. As envisaged,   we   reveal   that  $a$  and    $\beta$ have an  impact on  such shadow behaviors.  For specific values of $a$,   we  remark that the obtained  shadow shapes share certain similarities with the ones of the  Kerr black holes in  plasma   backgrounds.  Using the   Event Horizon Telescope  observational  data, we  provide   predictions for the stringy gravity  parameter   $\beta$   which could play a relevant role in M-theory compactifications. We finish  this  work by a discussion on the behaviors of  the light   rays  near  to   such   four dimensional    black holes by computing the  deflection angle in terms of  a required  moduli space.

	\end{abstract}
\newpage

\tableofcontents

\newpage
\section{Introduction}

The study of black hole physics  becomes a warm subject   approached from  different  gravity theories including the  modified ones.    This  relevant interest has been supported and  encouraged  by  the    observational    results  brought either  by  Event  Horizon Telescope (EHT)  collaborations    or   by   the gravitational wave detections  \cite{3,4,5,6}.   These physical activities are behind the most recent  development    dealing   with various  black hole properties  including the     thermodynamic and  the optical  ones.  These properties  have been  largely investigated by  considering  certain  gravity models living  in   different background   geometries such as  the Anti-Sitter space-times
(AdS)   \cite{R5,R6}. These non-trivial geometries open gates to approach  the  thermodynamic  aspect   by identifying       the cosmological constant with  the pressure \cite{19,Ref10,Ref7}.  This identification  has been explored    to investigate  the  thermodynamic   behaviors such as the  stability and   the  phase transitions of many black holes in  arbitrary dimensional gravity theories    including  supergravity ones like Type II  superstrings and M-theory scenarios\cite{18,181,182,Ref5,Ref6,17}.    In this context, several  thermodynamic quantities  have been  computed  and examined.  The entropy, the heat capacity and the Gibbs free energy  have been  calculated,  being    needed  to study local and  global stabilities. They have  been  exploited also in  the discussion of the 
 Hawking-Page transition between  a  large stable black hole  and a  thermal gas in AdS geometries  using analytical and numerical methods.  Using the result of  the AdS/CFT  conjecture,    the thermodynamics  and the  thermodynamical geometry of  the AdS black holes  from M-theory in the presence of
M2 and M5-branes with and without  dark energy  have been investigated\cite{180}.  Concretely,   the stability  and the phase transitions  of four and seven dimensional  AdS black holes have been studied in terms of the M-brane number.

Parallely,   the optical  properties of the  black holes     have been largely  investigated using analytical and numerical approaches. A close examination shows that  two relevant concepts have been  studied being  the shadow and the  deflection angle of  the light rays  near to   black holes  \cite{9,10,d1M,d2,d3H,d4H,d51,Belhaj2,Belhaj3, Carlo1,Carlo2}.   The first concept has been obtained via  the  Hamilton-Jacobi  algorithm  providing the equations of motion of photons  near  to a black hole solution \cite{K,BC}.  In four dimensions,  for instance, the shadow  behaviors have been approached in terms of  one-dimensional  closed real curves.  The   size and the shape   of such curves depend on the moduli space of the black hole  in question.   Precisely,      the non-rotating black hole involves perfect circular shadows  where the size could be controlled by certain parameters such as   the charge.   The spin  parameter,  however, affects such geometries   by providing  distortion deformations  leading to   non-trivial  configurations known as  D-shapes    \cite{Xa,J,RC,Belhaj4,Belhaj11}.   Among others, the shadows of   four-dimensional  AdS black holes in M-theory  scenarios   have been investigated in terms of the  M2-brane number. It has been shown that such  a number  controls  
the  geometric deformation of the shadows \cite{B12,ma}.    Using   the Gauss-Bonnet theorem,  moreover,   the deflection angle and the trajectory of the light rays  near   to black holes derived from  the compactification of M-theory     on the real spheres on $S^7$  and $ S^4$  have been studied \cite{hajar}. In particular,     the impact of the M-brane number and the rotating parameter  on such optical behaviors has been examined.

Recently,  modified gravity  models, being  used to solve  certain  Universe problems,  have been exploited to study  black holes. Many theories have been proposed using different roads and methods.  Precisely,  the most studied one is  Einstein Gauss-Bonnet (EGB) gravity   supported by string theory and related dual models including   M-theory  \cite{23Y,230Y,24}.   Black holes in such gravity models have been built and studied in terms of  a parameter called  Gauss-Bonnet  gravity parameter.  The thermodynamics and  the optical properties of these black holes  have been   dealt with  in terms of such a  parameter\cite{282Y,283Y}. It has been shown  that the  Gauss-Bonnet   parameter  has a relevant impact on the black hole physical  properties \cite{2855,285}. It has been observed that 
the shadow radius decreases by  increasing  this  parameter \cite{Sunny}.

More recently,   a   novel  Starobinsky-Bel-Robinson (SBR)  gravity    has been constructed by introducing a new  stringy parameter called $\beta$ \cite{SBR1}.    This four-dimensional gravity has been inspired by M-theory  described, at lower energies, by  eleven-dimensional supergravity.  Concretely, it has been suggested that this modified gravity could be derived  from the compactification  of M-theory on a  two sphere factor   being $S^3 \times S^4$ \cite{SBR1}.  Models in such a novel  gravity  have been proposed.  Concretely,  inflation scenarios   with the stingy  gravity parameter   $\beta$ have been  treated   where the associated  cosmological  observables  have been computed \cite{SBR2}. It has been suggested that  the  observational data can impose constraints  on such a parameter.  Moreover,  the  Schwarzschild-type black holes in   the  SBR  gravity have been built.   Precisely,    the corresponding  thermodynamic properties   have  been addressed   in terms of  $\beta$   by computing the relevant quantities including  the entropy and the pressure. An examination reveals that these quantities  are corrected  by the presence of $\beta$ \cite{SBR3}.   

The aim of  this work is to contribute to such activities by    investigating   the  optical properties of     black holes  in the  SBR   gravity  which could be embedded in M-theory scenarios.   Particularly,  we study   the shadows and the deflection angle of light rays by non-rotating and rotating  black holes  in four dimensions.   We start by discussing  the shadow  of the Schwarzschild-type solutions in such a novel gravity. As expected, we obtain  perfect circular configurations  where the size decreases with  the   stringy gravity  parameter $\beta$ being constrained by  the shadow existence.   Combining  the  Newman-Janis algorithm and  the Hamilton-Jacobi mechanism,  we   study the  shadow  behaviors of  the rotating solutions in terms of  one-dimensional real curves.   In the $\beta$ range,  we  find  various   sizes and shapes depending  on  the rotating and the gravity parameters $a$ and $\beta$, respectively.    To  examine  the shadow geometric deformations,   we  study   the astronomical observables and   the energy emission rate.  As  envisaged,   we   show  that the stringy  rotating parameter $a$  and   the gravity parameter $\beta$ have  a relevant impact on the shadow aspect. For specific values of $a$,   we   observe  that the derived   shadow shapes share certain similarities with the ones of the  Kerr black hole in  plasma   backgrounds  \cite{Z1,Z2}. Using the   observational  EHT data, we  provide  predictions for  $\beta$   which could play a  primordial role in the M-theory compactification.  Finally, we discuss  the  light  ray behaviors near  to the    SBR   black holes by calculating and analyzing  the deflection angle  in terms of the  rotating and the  stringy gravity  parameters. 

The organization of this paper is as follows.  In section  2, we reconsider the study of  the black holes in the   SBR  gravity.  In section 3, we investigate the shadow behaviors and provide  predictions  for  $\beta$ by the help of EHT data.  In section 4, we analyze  the light ray behaviors near to  the SBR  black holes  by computing  and analyzing the deflection angle variation. In the last section, we provide concluding remarks.

\section{Black holes in  SBR   gravity}
In this section, we give a concise reconsideration  on black hole solutions in  modified  gravity theories originated from higher dimensional space-times. In particular, we deal with certain black hole solutions in   the SBR   gravity which could be embedded  in M-theory living in  the  eleven dimensional space-time, recently reported in \cite{SBR1}.  It is  recalled that M-theory involves a specific bosonic sector containing a metric $g_{MN}$  and a tensor 3-form   $C_{MNP}$  coupled to M2-branes being dual to M5-branes \cite{MT}. The associated four-dimensional gravity models could be obtained using the compactification mechanism with the presence of stringy  fluxes required by the  stabilization scenarios \cite{SBR1,SBR3}. Roughly speaking, the action we consider here reads as 
\begin{equation}
S_{SBR}=\frac{M_{pl}}{2}\int d^4x  \sqrt{-g}\left(R+ \frac{R^2}{6m^2}- \frac{\beta}{32m^6}(P_4^2-E^2_4)\right)
\end{equation}
where $g$ is the determinant of the metric and $R$ is the Ricci curvature.   $m$    is a free  mass parameter which could have various interpretations depending  on the underlying theory. $\beta$  is a positive dimensionless coupling which will be  considered as  a relevant parameter in  the present investigation. It is worth noting that such a parameter depends on the M-theory compactification which could be  fixed  from  the black hole optical behaviors.  $P_4^2$ and $ E^2_4$ are quartic contributions associated with the Pontryagin and  the Euler topological    densities.
According to  \cite{SBR1,SBR3},  the  last term   is linked to the Bel-Robinson tensor  $T_{\mu\nu\lambda\rho}$ in four  dimensions  by means of the relation 
\begin{equation}
T^{\mu\nu\lambda\rho}T_{\mu\nu\lambda\rho}= \frac{1}{4}\left(P_4^2-E^2_4\right)
\end{equation}
where one has used 
\begin{equation}
T^{\mu\nu\lambda\sigma}=R^{\mu\rho\gamma \lambda}R^{\nu \sigma}_{\rho\gamma}+R^{\mu\rho\gamma\sigma}R^{\nu\lambda}_{\rho\gamma}-\frac{1}{2} g^{\mu\nu} R^{\rho\gamma\alpha\lambda}R_{\rho\gamma\alpha}^\sigma.
\end{equation}
 A close inspection reveals that the  SBR gravity  action involves two parameters $m$ and $\beta$.  This  action has been approached to provide  physical models including inflation \cite{SBR2}. Moreover,    the Schawarshild   type black holes  in such a gravity have been built where  the parameter $m$  has been absorbed  by solving the associated equation of motion   \cite{SBR3}. Concretely, the corresponding thermodynamic quantities   have been also computed and  investigated.  It has been shown that such quantities are corrected by the stringy gravity  parameter $\beta$.    In this way, the line element of this non-rotative solution  has been found to be 
\begin{eqnarray}
ds^2=-f(r)dt^2+\frac{1}{f(r)}dr^2+r^2d\Omega^2
\label{s1}
\end{eqnarray}
where the fundamental metric function $f(r)$ is given by
\begin{equation}
f(r)=1-\frac{r_s}{r}+\beta\left(\frac{ 4\sqrt{2}\pi G r_s}{r^3}\right)^3\left(\frac{108 r-97 r_S}{5r}\right)
\end{equation}
where one has used $ r_s=2GM$. Here, $G$  and $M$  represent the Newton constant and the mass parameter, receptively. Having considered the non-rotating case, we could  construct the rotating black holes in  the SBR gravity. To provide such  solutions, it  has been suggested an  useful algorithm  called  the Newmann-Janis algorithm \cite{JNA,JNABH}.  A close examination shows that this approach can be adopted for certain modified gravity models  by introducing extra parameters.
Supported by such activities, we would like to  investigate   rotating   black holes in the  SBR gravity.  Applying the Boyer-Lindquist coordinate systems, we propose  and  assume  the following metric line element 
\begin{eqnarray}
\label{aaa}
ds^2 &= &-\left(\frac{\Delta(r)-a^2\sin^2\theta}{\Sigma}\right)dt^2+\frac{\Sigma}{\Delta(r)}dr^2-2a\sin^2\theta\left(1-\frac{\Delta(r)-a^2\sin^2\theta}{\Sigma}\right)dtd\phi+\Sigma d\theta^2\nonumber \\
&+&\sin^2\theta\left[\Sigma+a^2\sin^2\theta\left(2-\frac{\Delta(r)-a^2\sin^2\theta}{\Sigma}\right)\right]d\phi^2\label{11}
\end{eqnarray}
where one has used 
\begin{eqnarray}
\Delta(r) & = & a^2+r^2 \left( 1-\frac{2 G M}{r}+\frac {1024 \pi ^3 \beta  G^6 M^3\left(108r -194 G M\right) }{5 r^{10}}\right)\\
\Sigma & =& r^2+a^2\cos^2\theta.
\end{eqnarray}
This line  element contains two parameters $a$ and $\beta$ recovering quite  known solutions.
Putting $a=0$, we  obtain  the previous non-rotating black holes. Taking $\beta=0$, however, we recover the Kerr metric  associated with the  following  delta function  
\begin{eqnarray}
\Delta_{Kerr}(r) & = & a^2+r^2 -2 G M r.
\end{eqnarray}
In what follows, the  stringy gravity  parameter $\beta$ will be crucial  to inspect the optical behaviors of  the  SBR  black holes going beyond the thermodynamic ones  reported  in \cite{SBR3}.   The aim of the  remaining part of this work is to investigate the  light   behaviors  around     the  SBR    black holes.  
\section{Shadows of SBR black holes}
The EHT  international collaboration has brought a black hole image which has been considered as  a relevant discovery in the underlying  physics  \cite{3,4}.  Indeed,  it  has been remarked that when the light passes near to a black hole, certain optical behaviors occur. Precisely, the light rays could be deviated quite strongly and could travel via circular geometrical  configurations. These considerations support the idea that a  black hole can be viewed as a dark disk in the sky. This disk is known by the shadow. Motivated by shadow activities, many black hole solutions have been investigated  using different approaches. These activities open gates to unveil more data on optical behaviors of black holes by  studying  the relevant concepts being the shadow and the deflection angle of  the light rays.  These concepts can be dealt with to provide  a test of    gravity model predictions  in the face of    the EHT observational data. This empiric exam could  fix certain gravity parameters originated from  M-theory compactifications. \\
In this section, we study  the shadow cast behavior  of  black holes  in such a  SBR gravity.  This study  will be made in terms of one-dimensional real curves  carrying the most data on the involved size and  the shape  deformations.  The present   optical behaviors can be approached via   the null geodesic equations of motion.   To derive such equations, we can  adopt   the Hamilton-Jacobi  algorithm  based on the following equation
\begin{equation}
\frac{dS}{d\tau}=-\frac{1}{2}g^{\mu \nu}\frac{d S}{dx^\mu}\frac{d S}{dx^\nu},
\end{equation}
where $\tau$ is the affine parameter.  $S$ represents the Jacobi action  which can be expressed as  follows
\begin{equation}
S=-E t+L \phi+S_{r}(r)+S_{\theta}(\theta),
\end{equation}
where $E$ and $L$ correspond to the energy and the momentum of the photon, respectively.  They are given by 
\begin{equation}
E=-p_t \qquad   L=-p_\phi.
\end{equation} 
$S_{r}$ and $S_{\theta}$ are functions of $r$ and $\theta$ variables, respectively. Using the separation method,  the four-dimensional   equations of motion  can be determined for  several black hole solutions including the  non-rotating and  the  rotating ones. 

\subsection{Shadows of non-rotating  SBR black holes}
\begin{figure*}[ht!]
		\begin{center}
		\begin{tikzpicture}[scale=0.2,text centered,<->]
		\hspace {-0.5cm}
\node[] at (0,0){\small  \includegraphics[scale=0.85]{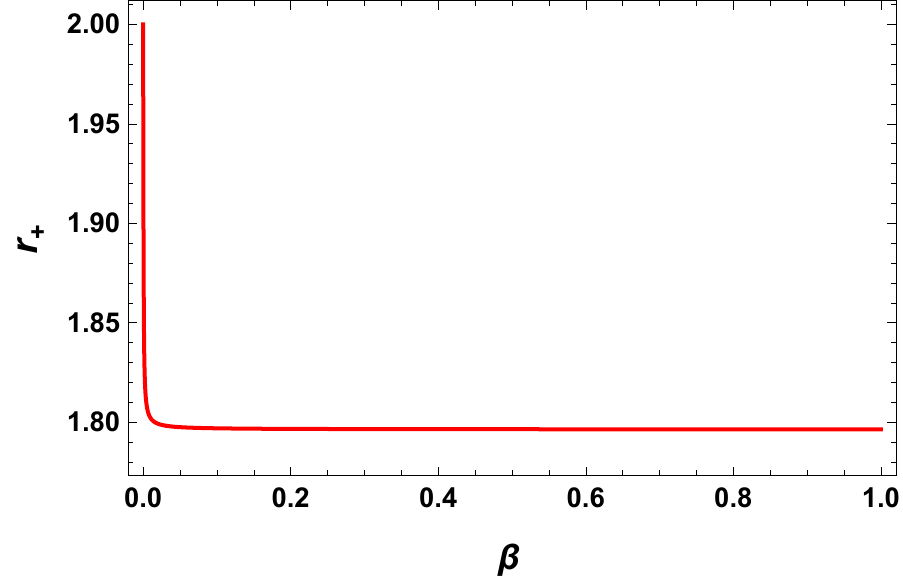}};
\node[draw, line width=0.9pt,color=blue,dashed] at (8,4.5){\small  \includegraphics[scale=0.4]{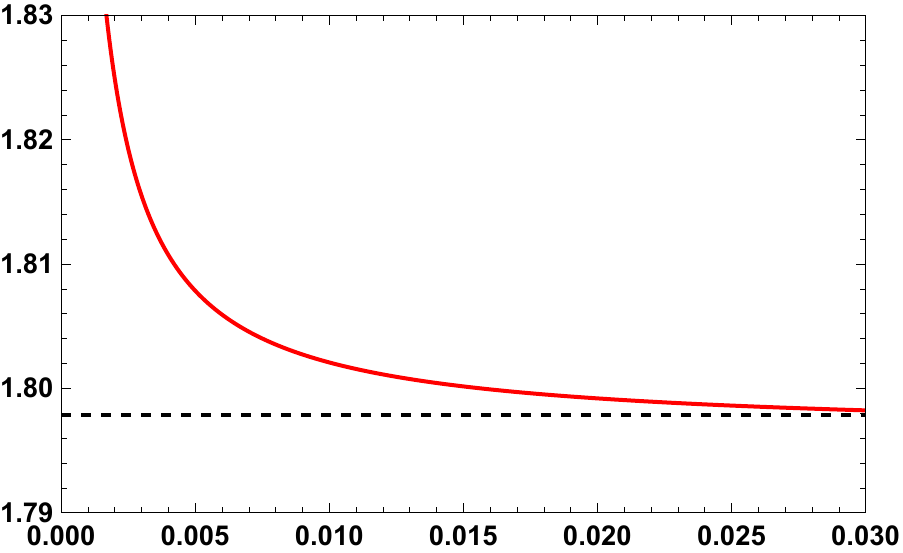}};
\node[draw, line width=0.9pt,color=blue,dashed] at (-12.7,-5.5){};
\node[] at (4.7,.5){{\fontsize{5}{5}\selectfont $r_+=1.79785$}};
\end{tikzpicture}
\caption{{\it \footnotesize Horizon size behaviors of SBR black holes by   varying the string parameter.}}
\label{f1}
\end{center}
\end{figure*}

\begin{figure}[!ht]
		\begin{center}
			
			\includegraphics[scale=0.7]{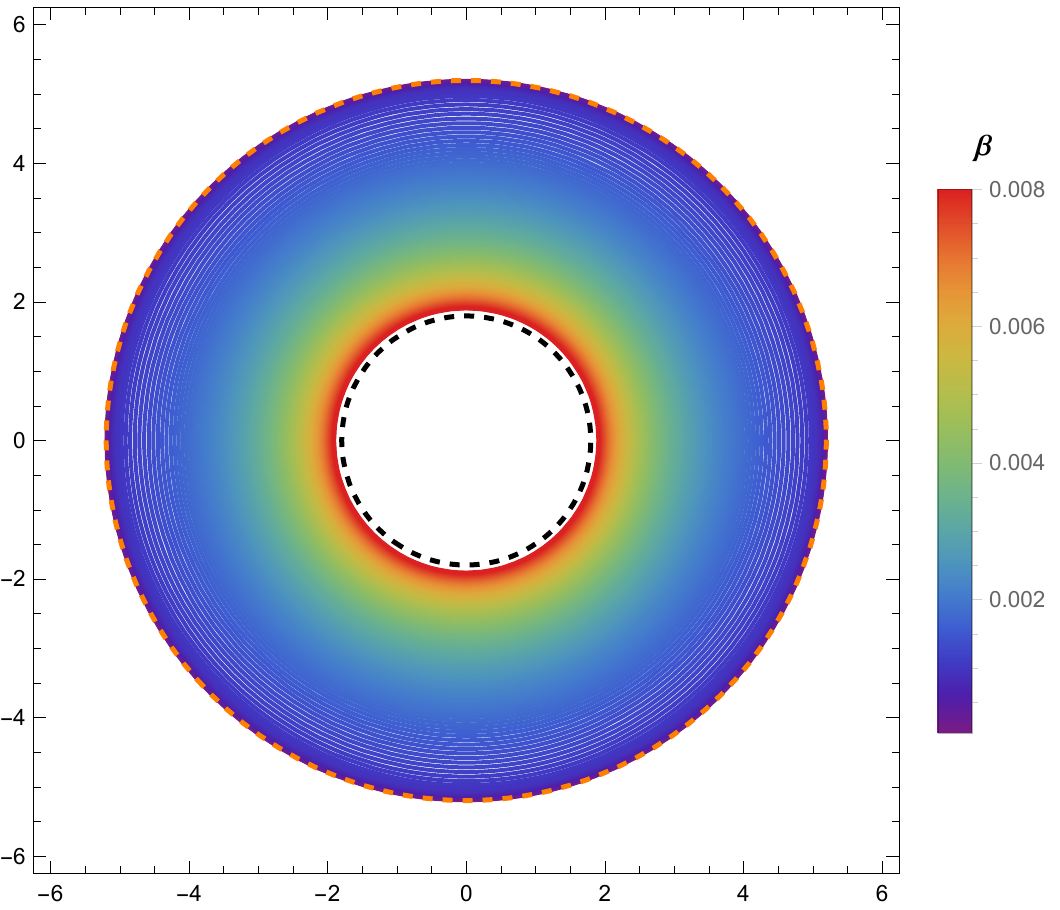} 
	 
\caption{{\it \footnotesize Shadow behaviors of SBR black by varying  the parameter  $\beta$. The orange and black  dashed circle represents the shadow of Schwarzschild and the horizon limite, respectively.}}
\label{xxxa}
\end{center}
\end{figure}

To get the associated shadows,  we  consider the black  hole solutions given by Eq.(\ref{s1}). Applying  the separation method \cite{K}, we   get the factorized relations
\begin{eqnarray}
 r^2f(r)\left( \frac{d S_{_{r}}(r)}{dr}\right) ^2-r^2\frac{E^2}{f(r)}+L^2&=&-\mathcal{K},\\
 \left( \frac{dS_{\theta}(\theta)}{d\theta}\right) +L^2\cot^2 \phi &=&\mathcal{K}, 
 \end{eqnarray}
 where $\mathcal{K}$ is the Carter constant which can be considered as a motion constant.  The equations of motion are
given by
\begin{eqnarray}
\frac{dt}{d\tau}&=&\frac{E}{f(r)}\\
r^{2}\frac{dr}{d\tau}&=& \pm \sqrt{-r^{2}f(r)\left(\mathcal{K}+L^{2}\right) +E^{2}r^{4}} \\
r^{2}\frac{d\theta}{d\tau}&=& \pm \sqrt{\mathcal{K}- L^{2}\cot^{2}\theta}\\
\frac{d\phi}{d\tau}&=&\frac{L}{r^{2}\sin ^{2}\theta}.
\end{eqnarray}
Using the radial equation, the effective potential takes the following form
\begin{eqnarray}
V_{eff}(r)=-\left( \frac{dr}{d\tau}\right) ^2. 
\end{eqnarray}
The calculation provides 
\begin{eqnarray}
V_{eff}(r)=\frac{f(r)}{r^{2}}\left(\mathcal{K}+L^{2}\right)-E^2.
\end{eqnarray}
This effective potential   should satisfy  the following conditions 
\begin{eqnarray}
V_{eff}(r)\Bigm|_{r=r_{c}}&=&0,\qquad \frac{dV_{eff}(r)}{dr}\Bigm|_{r=r_{c}}=0
\end{eqnarray}
where  $r_{c}$   represent the radius of the  unstable  circular orbits. This radius   can be  obtained by solving   the  equation
\begin{equation}
\label{x1x}
f'(r)-2f(r)=0
\end{equation}
where  the prime notation  denotes the derivative with respect   to the radial coordinate. A close examination shows that  the shadow geometries can be established  by means of   the  impact parameters. In four dimensions,  these parameters    read as  
\begin{equation}
\xi_{SBR}=\frac{L}{E} \qquad \eta_{SBR} =\frac{\mathcal{K}}{E^{2}}
\end{equation} 
which are given in  terms of the conserved quantities.  In the non-rotating case,  the computations give 
\begin{equation}
\xi_{SBR}^2+ \eta_{SBR} =\frac{25 r_c^{12}}{1390592 \pi ^3 \beta  G^7 M^4-663552 \pi ^3 \beta  G^6 M^3 r_c-20 G M r_c^9+15 r_c^{10}}
\label{sa}
\end{equation}
 generating one-dimensional real curves in a two-dimensional plane.  To get a more clear configurations,   the celestial coordinates should be exploited. Indeed, they are given by 
\begin{eqnarray}
\small
X_{SBR} &=& \lim_{r_{_{o}\rightarrow +\infty}} \left( -r^2_{o} \sin \theta_{o} \frac{d\phi}{d r}\right) \\
Y_{SBR} &=&  \lim_{r_{_{o}\rightarrow +\infty}}\left( r^2_{o}\frac{d\theta}{dr}\right) 
\end{eqnarray}
where $r_{o}$  indicates  the distance of the observer from the black hole, while  $\theta_{o}$  represents  the angle of  the inclination between the observer line of  sight and the axis of the black hole rotation.  In this way, the  left-hand side of Eq.(\ref{sa})  can be rewritten as  follows 
\begin{equation} 
\xi^2_{SBR}+\eta_{SBR}=X^2_{SBR}+Y^2_{SBR}.
\end{equation}
To get the shadow cast representation,  we should handle this equation  by combining  Eq.(\ref{x1x}) and  Eq.(\ref{sa}) via a numerical method.   Before going ahead, we first  inspect the   event horizon radius $r_+$ behaviors in terms of the stringy  gravity  parameter $\beta$.   Fig.(\ref{f1})   gives  the horizon radius   as a  function of  such a  parameter. 
In certain regions of the moduli space,  it  follows from  this figure  that the horizon radius of the  SBR black holes decreases by increasing  $\beta$.  For values  $\beta\geqslant3.10^{-2}$,  however,  the horizon radius  keeps a  constant  value around $r_+\simeq1.798$.
 Taking into account such constraints,  we can now investigate    the shadow behaviors  of  the non-rotating black holes in  the SBR    gravity  by varying the  stringy gravity   parameter and fixing the above ones namely $G$  and $M$    to  1.  Fig.(\ref{xxxa})  depicts the shadows of  the SBR black holes  as  a  function of  $\beta$.
  As expected,  the  obtained shadow geometries   are  perfect circles.  We  observe that  the stringy   gravity parameter $ \beta$ controls  their  sizes.  Precisely, they    decrease  by increasing  $\beta$.      A close examination on shadow behaviors shows that  we should impose extra conditions  on   $\beta$.  For  $\beta>8.10^{-2}$, we remark that  the radius of the horizon becomes  large that the shadow  one. This leads to  a condition on the shadow existence  matching  perfectly with the findings associated with  the cosmological  building models    embedded in the   SBR   gravity  \cite{SBR2}.  These  small values  of $\beta$ can be supported by others investigations \cite{SBR1,SBR3}. This could  provide certain bridges between  inflation and black holes built  from  the SBR    gravity. Having investigated the  non-rotating case, we move to consider the implementation of  the rotating parameter    in   the shadow discussions.  
\subsection{ Shadow  behaviors of rotating SBR black holes }
In this section, we would like to  study  optical   behaviors of the  rotating black holes in  the SBR  gravity.   Precisely, we investigate  the shadow behaviors in such M-theory inspired black hole models.
\subsubsection{Shadows of rotating SBR black holes }
Here, we consider the rotating version of  the black holes  in  the SBR  gravity.  Using the separation method to the metric (\ref{aaa}), we get
\begin{eqnarray}
\Sigma\frac{dt}{d\tau} & = & \frac{r^2+a^2}{\Delta}\left(E(r^2+a^2)-aL\right)-a\left(aE\sin^2\theta-L\right), \\
\Sigma\frac{dr}{d\tau} & = & \pm\sqrt{\mathcal{R}(r)},\\
\Sigma\frac{d\theta}{d\tau} & = & \pm\sqrt{\Theta(\theta)},\\
\Sigma\frac{d\phi}{d\tau} & = & \frac{a}{\Delta}\left(E(r^2+a^2)-aL\right)-\left(aE-\frac{L}{\sin^2\theta}\right),
\end{eqnarray}
where $\mathcal{R}(r)$ and $\Theta(\theta)$ read as
\begin{eqnarray}
\mathcal{R}(r) & = & \left((r^2+a^2)E-aL\right)^2-\Delta\left((aE-L)^2+\mathcal{K}\right) , \\
\Theta(\theta)  &= & \mathcal{K}-\left(\frac{L^2}{\sin^2\theta}-a^2E^2\right)\cos^2\theta.
\end{eqnarray}
The unstable photon orbits  should satisfy the constraints
\begin{equation}
\mathcal{R}(r) \bigg\rvert_{r=r_0}= 0, \qquad \frac{\partial\mathcal{R}(r)}{\partial r} \bigg\rvert_{r=r_0}=0.
\end{equation}
 In the case of  the rotating solutions, the impact parameters  take the following forms 
\begin{eqnarray}
\eta_{SBR} & = & \frac{r^2\left[16\Delta(r)(a^2-\Delta(r))-r^2\Delta '(r)^2+8r\Delta(r)\Delta '(r)\right]}{a^2\Delta '(r)^2}\bigg\rvert_{r=r_0},\\
\xi_{SBR} & =& \frac{(a^2+r^2)\Delta '(r)-4 r\Delta(r)}{a\Delta '(r)}\bigg\rvert_{r=r_0}.
\end{eqnarray}
Using the fact that $\beta$ is very small,    the computations give
 \begin{eqnarray}
\eta_{SBR} &=& \eta_{KR}+\beta\gamma_1\\
\xi_{SBR}& =&\xi_{KR}+\beta\gamma_2  
\end{eqnarray}
where  $\eta_{KR}$ and $\xi_{KR}$  are  the impact parameters of  the Kerr black hole which read as 
\begin{eqnarray}
 \eta_{KR}&=&\frac{r_0^3 \left(4 a^2 G M-r_0 (r_0-3 G M)^2\right)}{a^2 (r_0-G M)^2}\\
\xi_{KR} & =&\frac{r_0^2\left(3 G M -r_0\right)-a^2\left(G M+r_0\right)}{a\left(r_0-GM\right)}.
\end{eqnarray}
  $\gamma_1$ and $\gamma_2$  are two  independent functions  of  $\beta$  given by 
\begin{eqnarray}
 \gamma_1&=&8192 \pi ^3 G^6 M^3\Big[\frac{ \left(a^2 \left(1261 G^2 M^2-1106 G M r_0+243 r_0^2\right)\right)}{5 a^2 r_0^6 (r_0-G M)^3}\nonumber\\
 &+&\frac{r_0 \left(-2619 G^3 M^3+3624 G^2 M^2 r_0-1646 G M r_0^2+243 r_0^3\right)}{5 a^2 r_0^6 (r_0-G M)^3}\Big] \\
\gamma_2& =&\frac{4096 \pi ^3 G^6 M^3 \left(a^2 (388 G M-189 r_0)+r_0\left(-873 G^2 M^2+917 G M r_0-243 r_0^2\right)\right)}{5 a r_0^8 (r_0-G M)^2}.\nonumber
\end{eqnarray}
It is denoted that  the  impact parameters of the  SBR black holes depend on $\beta$   and  the ones of the Kerr black hole.  For $\beta=0$,  such impact  parameters  reduce to the Kerr ones. As envisaged,     $\beta$  could  affect the geometry of  the shadows by means of the size and the shape.  According to   \cite{K,BC},   the coordinates of a celestial plane $(X_{SBR},Y_{SBR})$   representing   all projections of  the spherical photon orbits will be exploited  in the computations. Indeed, they are  given by 
\begin{eqnarray}
X_{SBR}& =& -r_o\frac{p^{(\phi)}}{p^{(t)}}=\lim_{r_o\to\infty}\left(-r_o^2\sin\theta_o\frac{d\phi}{dr}\right), \\
Y_{SBR} & =& r_o\frac{p^{(\theta)}}{p^{(t)}}=\lim_{r_o\to\infty}(r_o^2\frac{d\theta}{dr})
\end{eqnarray}
where $r_o$ and $ \theta_o$ denote  the distance and the angle of the observer, respectively.  Such    celestial coordinates and the  above  impact parameters are  linked via   the following relations  
 \begin{eqnarray}
X_{SBR} & = & -\xi_{SBR}\csc\theta_0,\\
Y_{SBR} & = & \pm\sqrt{\eta_{SBR}+a^2\cos^2\theta_0-\xi_{SBR}^2\cot^2\theta_0}. 
\end{eqnarray}
In the equatorial plane $\theta = \pi/2$, $X_{SBR}$ and $Y_{SBR}$ reduce to
\begin{eqnarray}
X_{SBR} & = & -\xi_{KR}+\beta\gamma_2,  \\
Y_{SBR} & = & \pm\sqrt{ \eta_{KR}+\beta\gamma_1}.
\end{eqnarray}
Vanishing $\beta$,  we recover the  Kerr situation. 
\begin{figure*}[!ht]
		\begin{center}
		\begin{tikzpicture}[scale=0.2,text centered]
		\hspace{0 cm}
\node[] at (-35,1){\small  \includegraphics[scale=0.60]{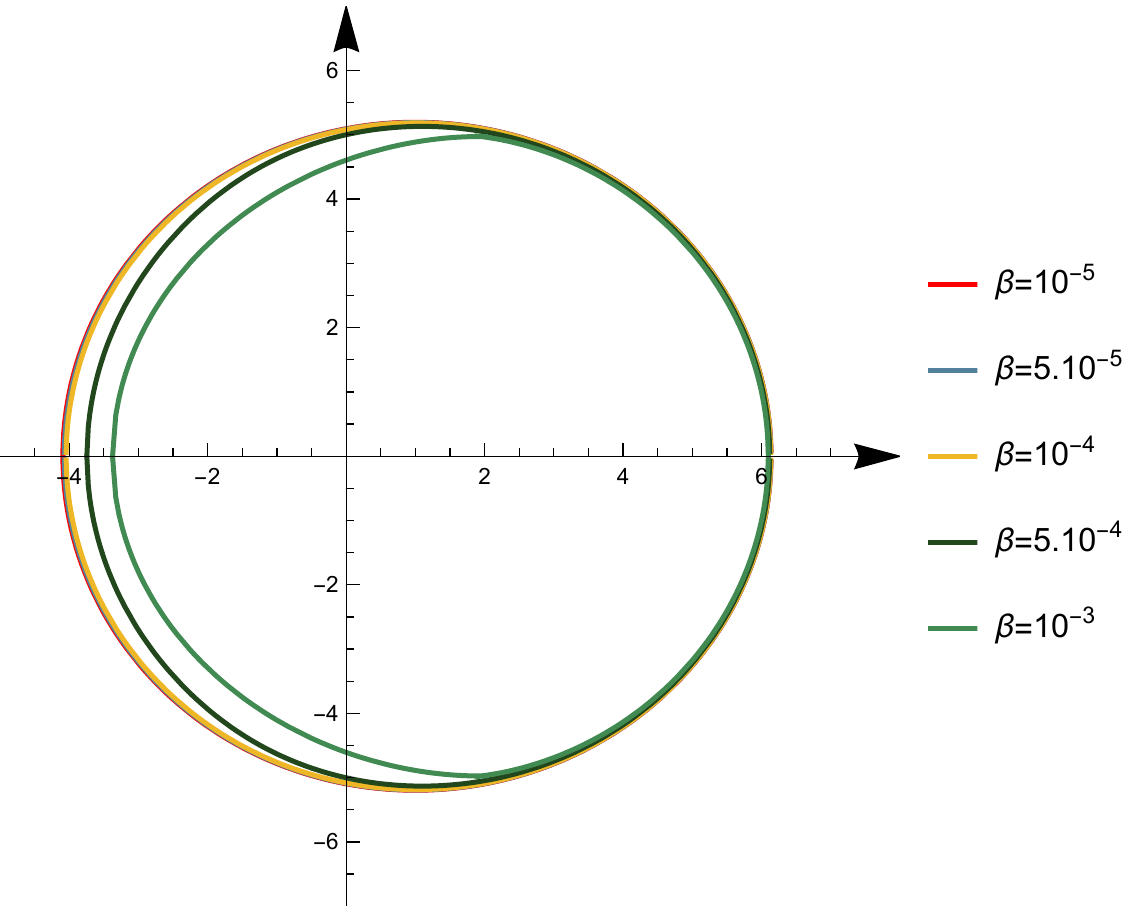}};
\node[] at (10,1){\small  \includegraphics[scale=0.60]{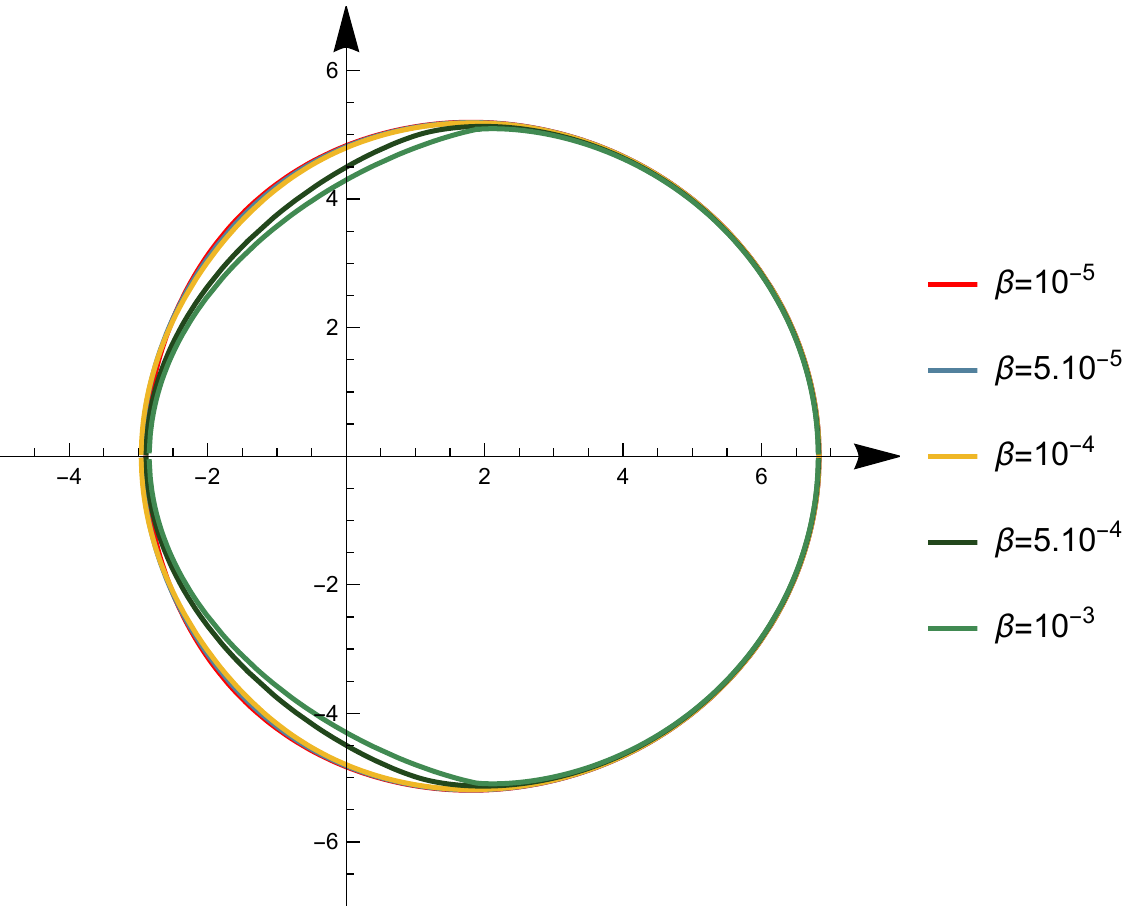}};
\node[] at (-36,-35){\small  \includegraphics[scale=0.60]{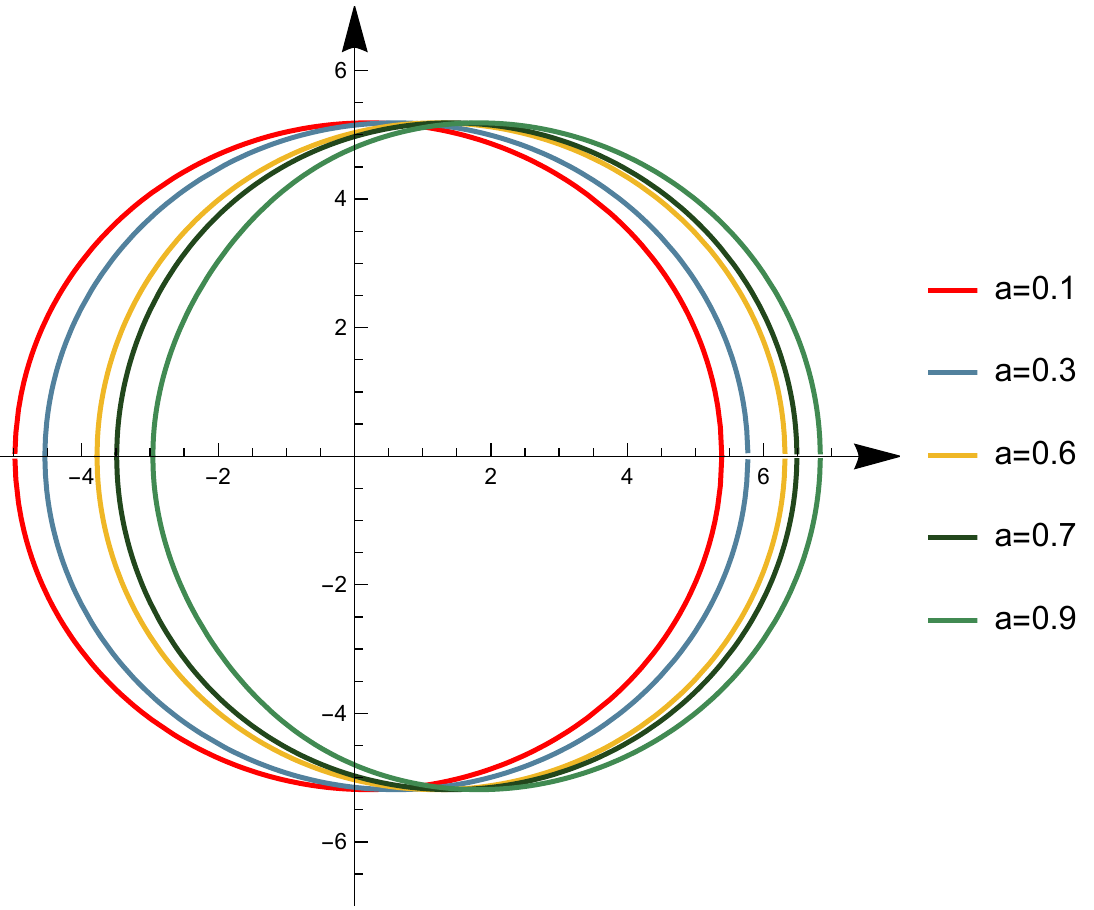}};
\node[] at (9,-35){\small  \includegraphics[scale=0.60]{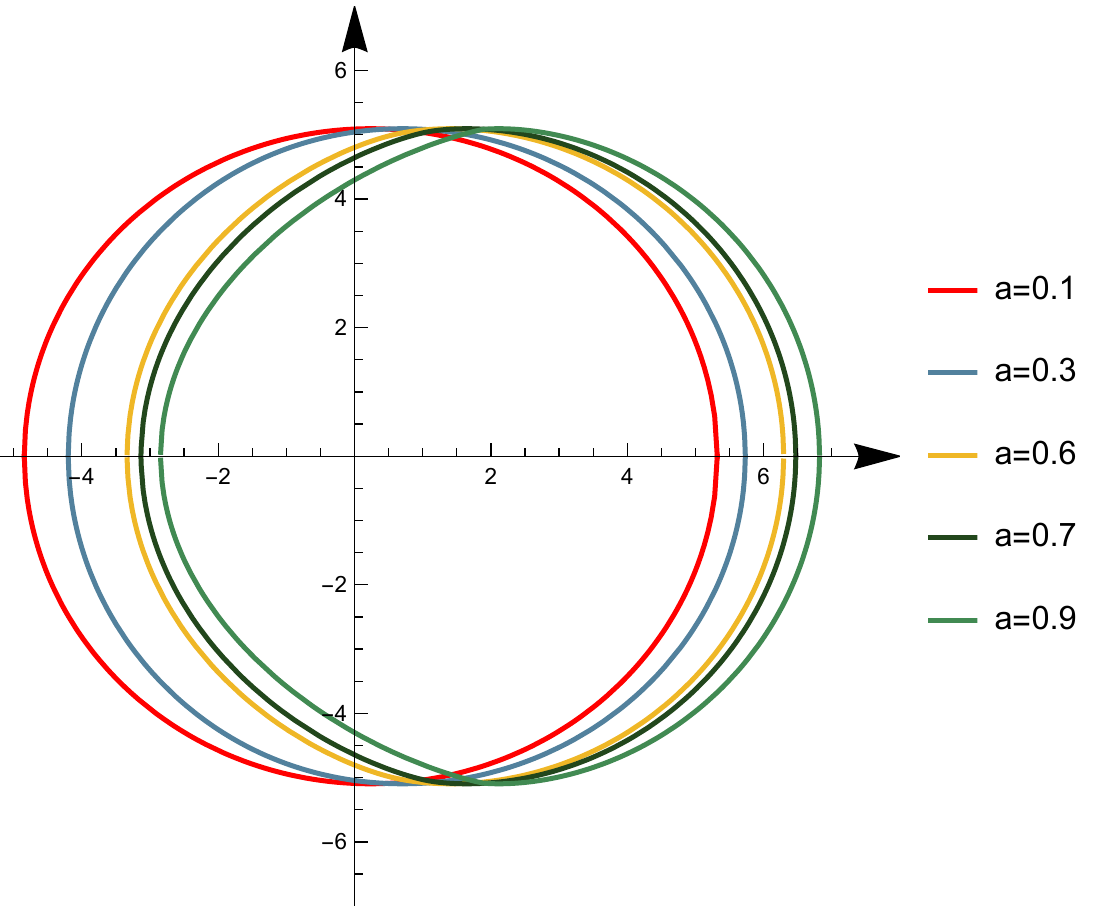}};
\node[color=black] at (-40,18.5) {$a = 0.5 $};
\node[color=black] at (5,18.5) {$a = 0.9$};
\node[color=black] at (-39,-17) {$\beta =9.10^{-5} $};
\node[color=black] at (6,-17) {$\beta = 7.10^{-4} $};
\node[color=black,rotate=90] at (-43.5,13) {\tiny{$Y_{SBR} $}};
\node[color=black,rotate=90] at (-53.5,0.7) {\tiny{$X_{SBR} $}};
\node[color=black,rotate=90] at (1.5,13) {\tiny{$Y_{SBR} $}};
\node[color=black,rotate=90] at (-8.5,0.7) {\tiny{$X_{SBR} $}};
\node[color=black,rotate=90] at (-43.5,-23) {\tiny{$Y_{SBR} $}};
\node[color=black,rotate=90] at (-53.5,-35) {\tiny{$X_{SBR} $}};
\node[color=black,rotate=90] at (1.5,-23) {\tiny{$Y_{SBR} $}};
\node[color=black,rotate=90] at (-8.5,-35) {\tiny{$X_{SBR} $}};

\end{tikzpicture}	
\caption{{\it \footnotesize Shadow shapes  of SBR black holes in the equatorial plane by considering  different   values of $a$ and $\beta$.}}
\label{F22}
\end{center}
\end{figure*}
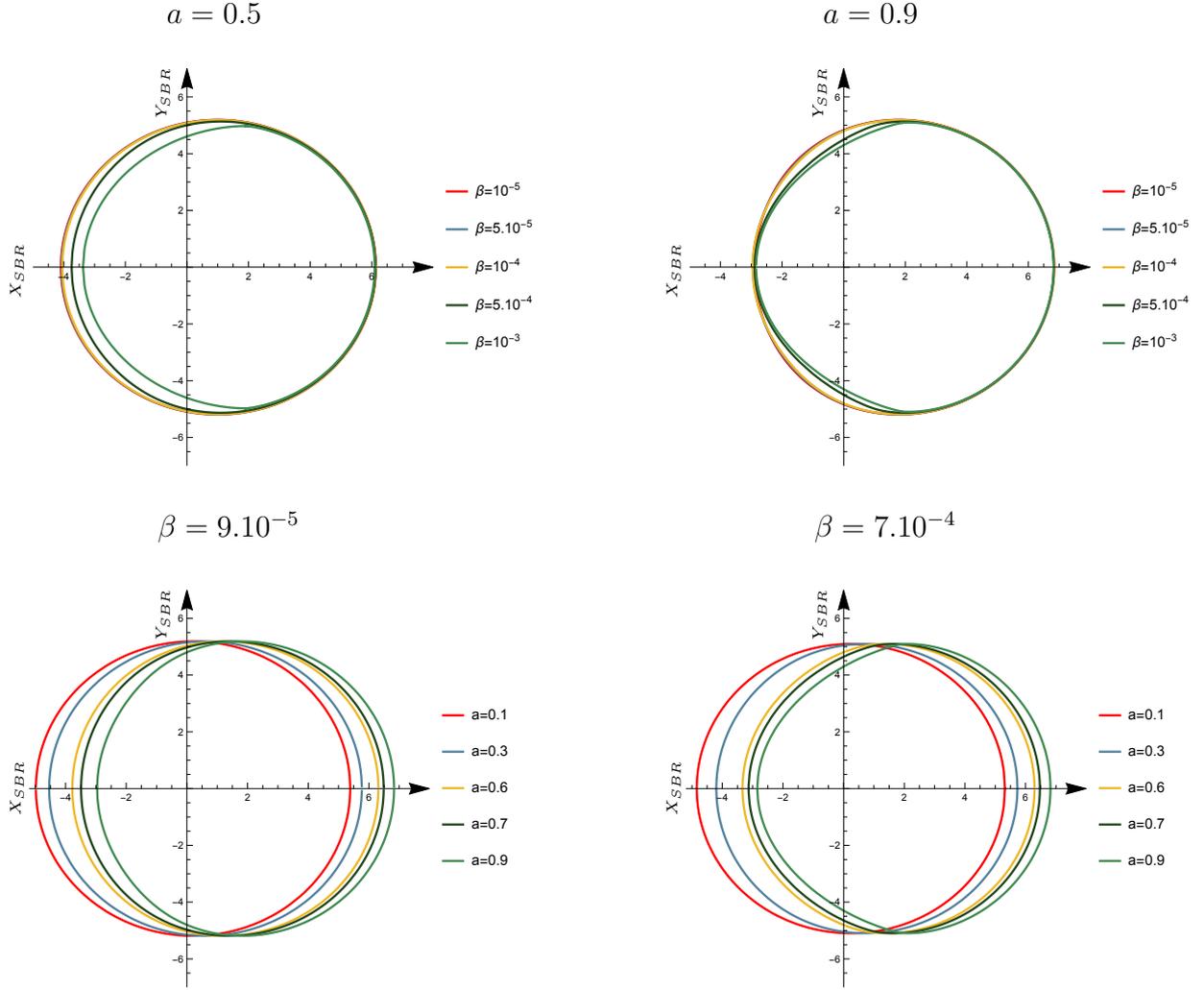
In Fig.(\ref{F22}), we  plot    the shadow behaviors  for certain values of the involved parameters including the stringy gravity parameter  $\beta$.  This figure  provides   various  shadow  sizes and shapes for different values of $a$ and $\beta$.  For fixed values of the rotating parameter,  we observe  that the shadows of  the SBR black holes increase  by decreasing  $\beta$. However,  the  geometry deformation  of  the SBR black holes  becomes relevant with $\beta$. For large values of $a$ and $\beta$, we find special shadow shapes analogue to  the ones appearing   in the case of   the Kerr black hole  in the presence of  Gaussian plasma distributions\cite{Z1,Z2}.  We would anticipate that  the plasma and the stringy gravity parameter involve similar influences on  the shadow shape of the rotating black holes. These behaviors appear for large values of $a$ and $\beta$. To check  that, we  consider two values of   $\beta$ namely  $\beta=9.10^{-5}$ and  $\beta=7.10^{-4}$. For the first value,  the shape deformations   remain negligible  by varying the rotating parameter. Taking  $\beta=7.10^{-4}$, however,  the   shape deformation becomes relevant by augmenting  such a  parameter.
\subsubsection{ Geometric deformations of the   shadow}
\begin{figure}[ht!]
\begin{center}
\includegraphics[scale=0.5]{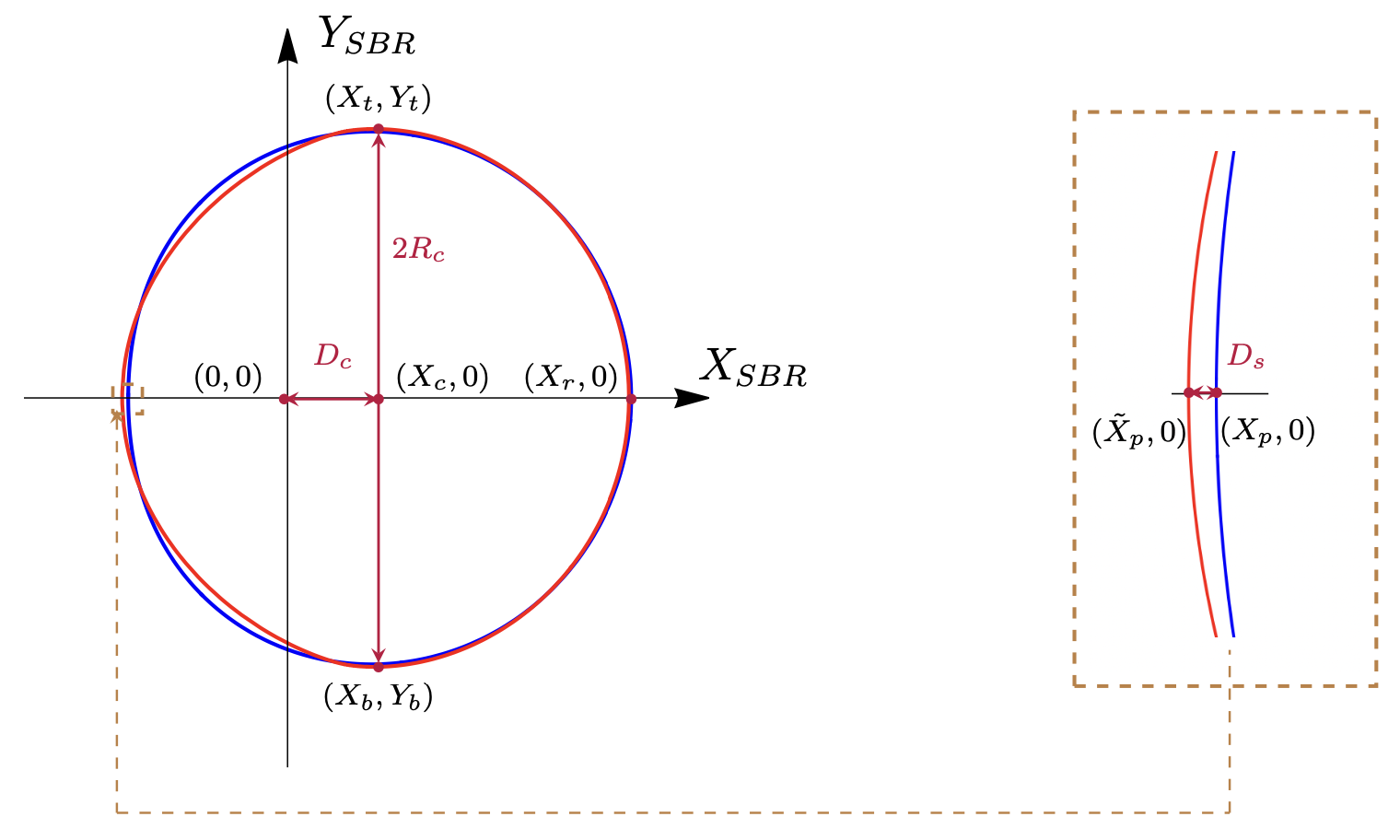}
\caption{{\it \footnotesize Illustration of  shadows size and shape  parameter of SBR black hole. The reference circle is represent by the blue color  and the red geometry is associated with the shadow of SBR black hole.}}
\label{ben1}
\end{center}
\end{figure}

\begin{figure*}[!ht]
		\begin{center}
		\begin{tikzpicture}[scale=0.2,text centered,<->]
		\hspace {-0.5cm}
\node[] at (-38,1){\small  \includegraphics[scale=0.520]{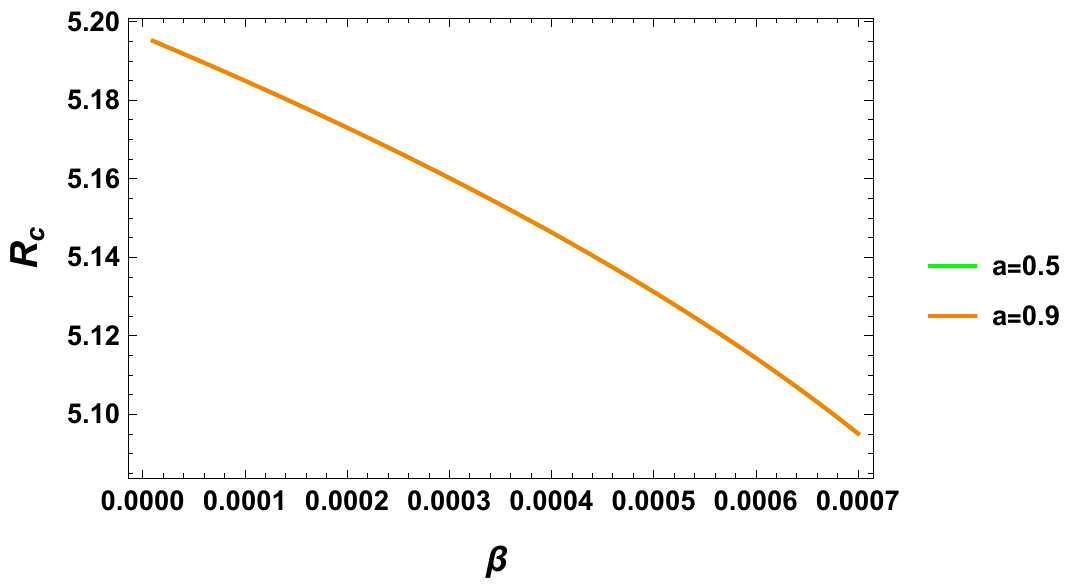}};
\node[] at (-9,1){\small  \includegraphics[scale=0.520]{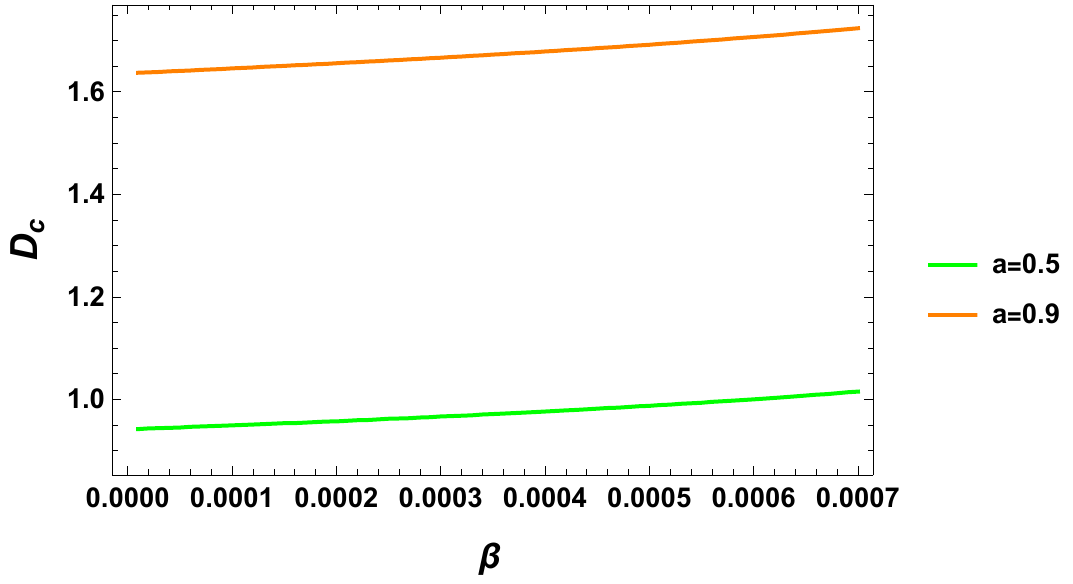}};
\node[] at (20,1){\small  \includegraphics[scale=0.520]{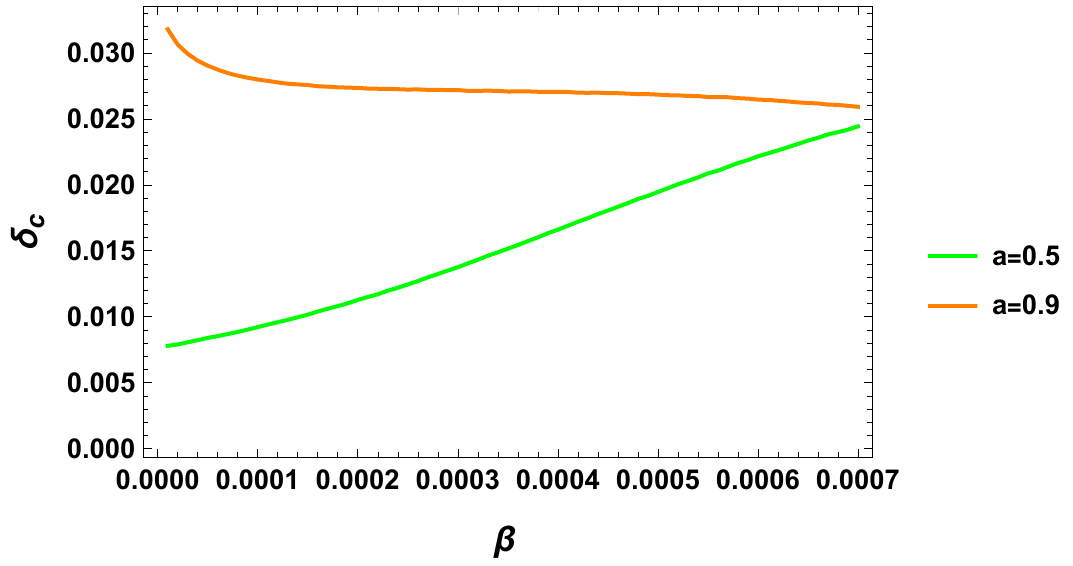}};
\node[] at (-37.5,-20){\small  \includegraphics[scale=0.520]{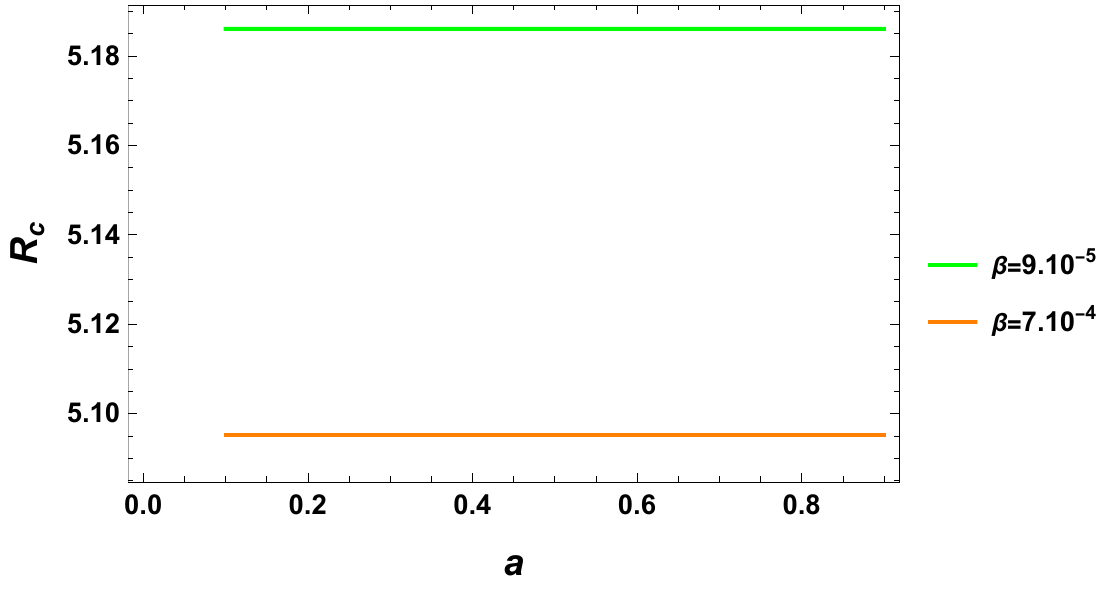}};
\node[] at (-8.5,-20){\small  \includegraphics[scale=0.520]{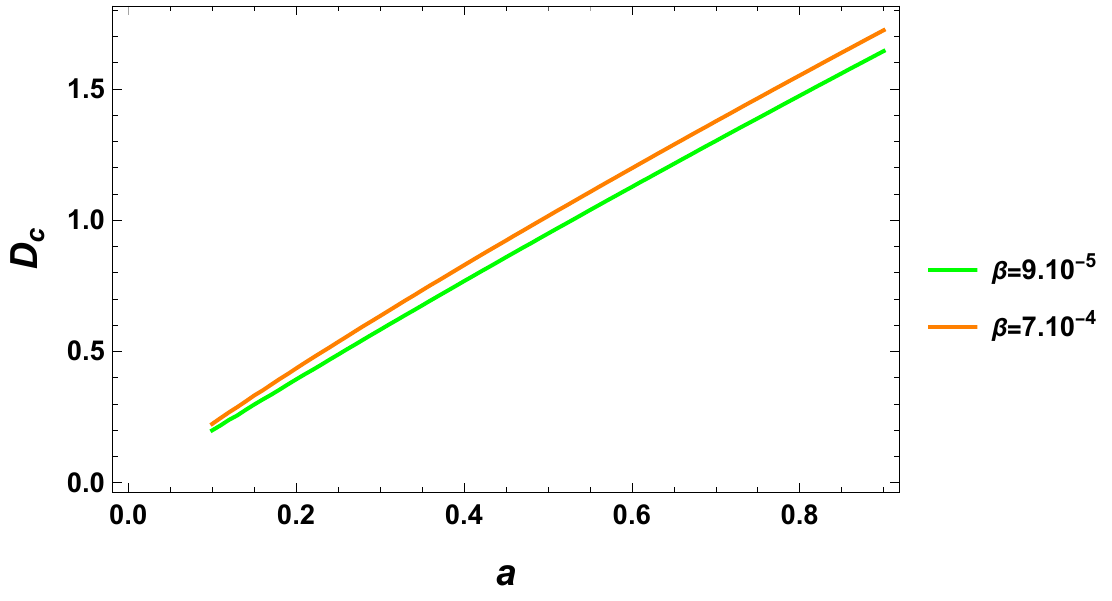}};
\node[] at (20.5,-20){\small  \includegraphics[scale=0.520]{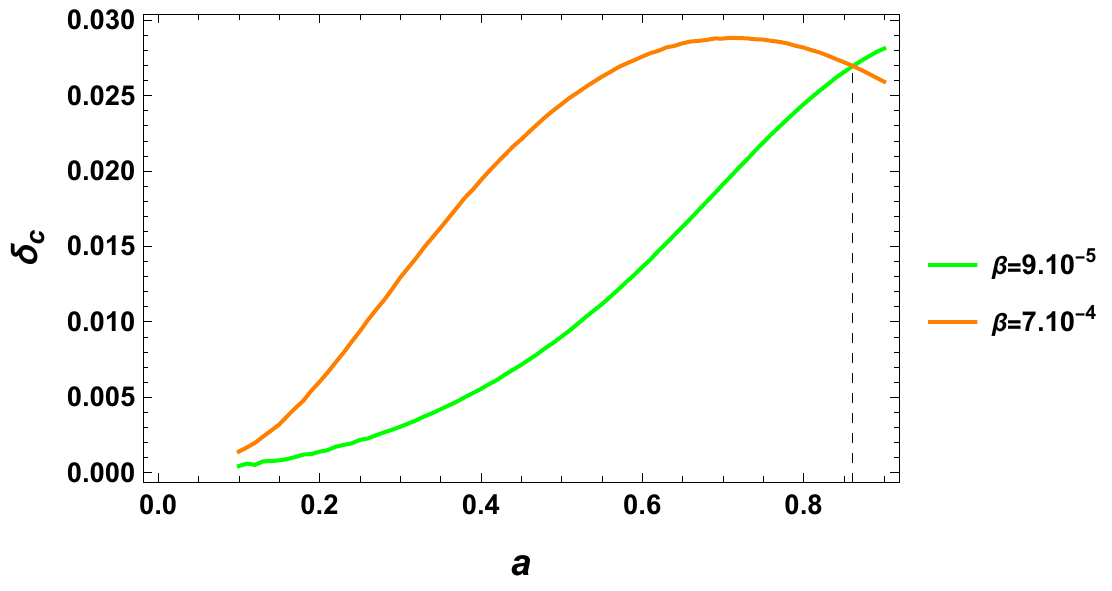}};
\node[rotate=90] at (27.5,-20){{\fontsize{5}{5}\selectfont $a=0.861$}};

\end{tikzpicture}
\caption{{\it \footnotesize Deformation observation quantities  of SBR black hole as function of the  rotating parameter and the string parameter.}}
\label{ben2}
\end{center}
\end{figure*}

The non-trivial   shadow geometries  of  the SBR black holes motivate one to inspect    the associated  deformations.   The geometric  deformations usually    can be  approached  in terms of   two parameters namely   $R_c$ and  $\delta_c$ controlling the size and the shape, respectively.   In order to   examine  such parameters,   Fig.(\ref{ben1})  illustrates  the shadow  deformations   in terms of a reference circle represented by  the blue color.
It  is recalled that   $R_c$   denotes  the shadow  maximal radius  passing through    three extreme points.  According to \cite{RC,B12,Belhaj11}, two ones   are  given by   the upper and the  lower positions of the  shadow  being $(X_t,Y_t)$ and  $(X_b,Y_b)$, respectively.   However,  the remaining one is  identified with  either   the reference circle $(X_p,0)$ or  $(\tilde{X_p},0)$  of the   shadow geometry.  In the perfect circular geometries, they coincide.  The displacement  $D_c$ of the  shadow from the centre  can be  obtained   by  computing 
\begin{equation}
D_c=X_r-R_c
\end{equation}
where  $R_c$  is the shadow  maximal  radius which could be  computed  from the relation
\begin{equation}
\label{ }
R_c=\frac{Y_t-Y_b}{2}.
\end{equation}
In this way,  the shape parameter  $\delta_s$   can be expressed as follows
\begin{equation}
\delta_c=\frac{D_s}{R_c}=\frac{|\tilde{X}_p-X_p|}{R_c}.
\end{equation}
 To  examine   the geometric deformations,   we   approach  $R_{c}$,   $D_c$  and   $\delta_{c}$  by varying the   rotating and   the stringy  gravity parameters.  Fig.(\ref{ben2}) illustrates the corresponding  computations.   
Fixing  the rotating parameter,   the shadow size decreases  by increasing  $\beta$. However,  the displacement of the shadow centre   and the  distortion deformation    increases with  the stringy  gravity  parameter.  It has been remarked that  $\delta_{c}$  takes higher values for a large rotating parameter. Fixing  the  stringy gravity  parameter $\beta$, the shadow size   of    the SBR black holes   remains almost constant   even we vary    $a$.  It has been observed   that the right displacement of the shadow  increases  with $a$   via  a linear behavior.    In 	addition,  we remark that the  shadow shape  deformation becomes relevant by   increasing the  rotating parameter.  Taking $\beta=7.10^{-4}$, for instance, the values of the deformation parameter  takes  values bigger than  the case of   $\beta=9.10^{-5}$.  A close examination shows that,  for  $a=0.861$,   a critical behavior where the curves intersect. For $\beta=7.10^{-4}$, this intersecting point provides a  changing of  such  a  distortion  behavior.  This  change could be  explained  by  the existence of   a    pair of cusps  in the shadow shape which matches  with the previous  shadow illustrations.  For a  specific range  of $\beta$,  we remark certain similarities  between the shadow sizes of  the SBR black holes  and  the Kerr black hole ones \cite{SBR1,SBR2,SBR3}.   Precisely, the shape  deformation of  the SBR black holes  could be compared  with the case of the  Kerr black hole in   plasma  backgrounds\cite{Z1,Z2}. This observation may  deserve a concrete  investigation in future works.

\subsubsection{ Energy emission rate}
\begin{figure}[!ht]
		\begin{center}
		\centering
			\begin{tabbing}
			\centering
			\hspace{-1.cm}\=\kill
			\includegraphics[scale=0.7]{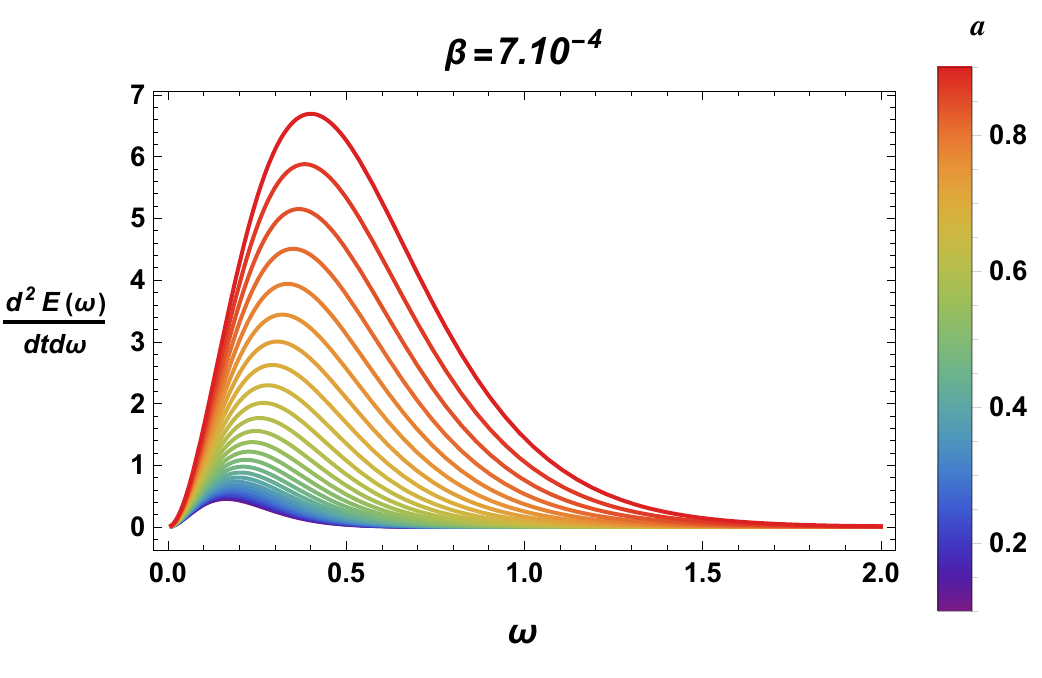} 
	\hspace{1cm}		\includegraphics[scale=0.7]{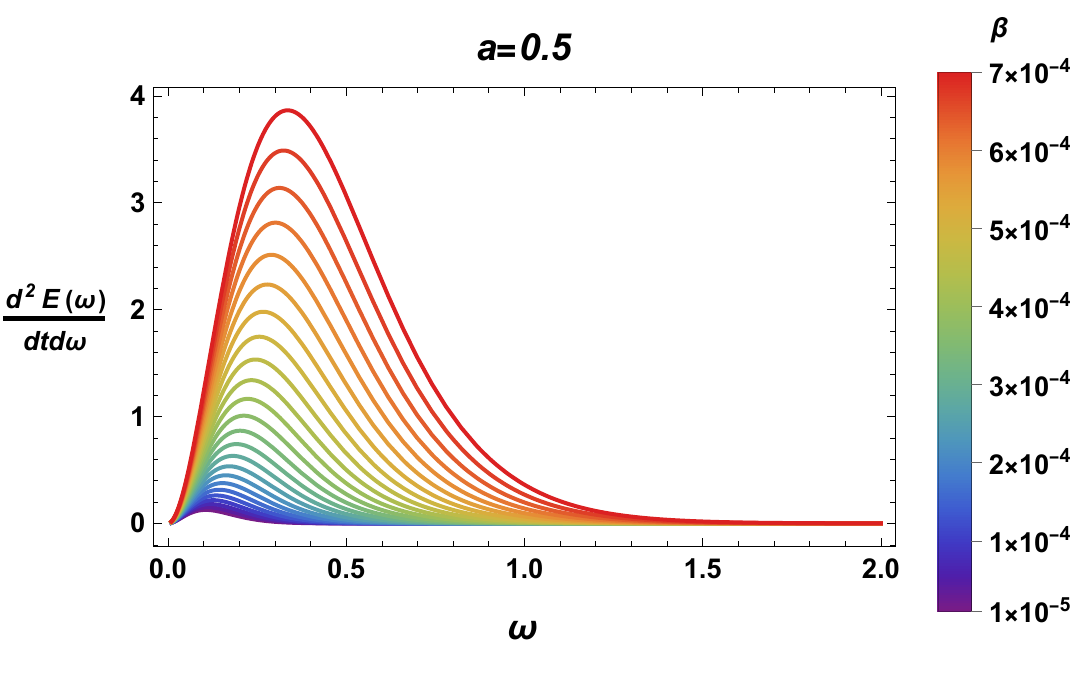}\\ 
	    \end{tabbing}
\caption{{\it \footnotesize Energy emission rate behaviors  of SBR black hole by varying $a$ and $\beta$ parameters.}}
\label{xxx}
\end{center}
\end{figure}

To complete the  discussion     of the shadow properties of    the SBR  black holes, we  study     the associated  energy emission rate   by considering   appropriate  approximations \cite{em}.   In the limit of   the high energies with   a far   distant observer,    the absorption cross-section   is  considered  as a geometrical optical limit   corresponding to   the black hole  shadows.   In fact,   this could  oscillate around   this   geometrical limit approached   in terms of the geodesic behaviors.  Following  \cite{em},    the energy emission rate  reads as 
\begin{equation}
\frac{d^2E(\omega)}{d\omega dt}=\frac{2\pi^3 R_c^2}{e^{\frac{\omega}{T_h}-1}}\omega^3
\end{equation}
where    $T_h$   and    $E(\omega)$  represent  the Hawking temperature and   the energy of the black hole as functions of  the frequency  $\omega$,  respectively.  $R_c$   denotes  the   shadow radius.    In the present work,  the  Hawking temperature can be obtained via the relation 
\begin{equation}
T_h=\frac{\Delta'(r)}{4 \pi (r^2+a^2)}\Big\vert_{r=r_{h}}
\end{equation}
where $r_h$   is  the event horizon radius.  This temperature  is found to be 
\begin{equation}
T_h=\frac{2 \left(794624 \pi ^3 \beta  G^7 M^4-387072 \pi ^3 \beta  G^6 M^3 r-5 G M r^9+5 r^{10}\right)}{5 r^9 \left(a^2+r^2\right)}\Big\vert_{r=r_{h}}.
\end{equation}
As usually, we study the  energy emission rate   behaviors  as a  function of the stringy   gravity parameter  $\beta$.   Fig.(\ref{xxx}) illustrates    the variation of the  energy emission rate   with respect to the  frequency.  
In the  left  panel of the figure, we  discuss   the rotating parameter effect  by fixing   $\beta$.  It follows that   the  energy emission rate  augments   with   the rotating parameter.   We  observe that higher values  of  such  energies correspond to large values of $a$.    Taking $a=0.5$,  we  discuss  the effect of $\beta$   in the  right  panel. Precisely, we remark that  the energy emission rate increases by increasing   $\beta$.   A close examination shows that the evaporation scenario  of  the SBR black holes is  fast compared to the  Kerr black hole and the non-rotating ones\cite{B12,d1M,Belhaj5,Belhaj1}.  Moreover, we could anticipate that  the  energy emission rate  of  the SBR black holes could involve similarities with the case of the  Kerr black holes  in the presence of plasma\cite{Atamurotov1,Babar1}.

\subsubsection{Constraints on the stringy  gravity parameter via EHT  observational   data}
\begin{figure*}[!ht]
		\begin{center}
		\begin{tikzpicture}[scale=0.2,text centered]
		\hspace{0 cm}
\node[] at (-35,1){\small  \includegraphics[scale=0.60]{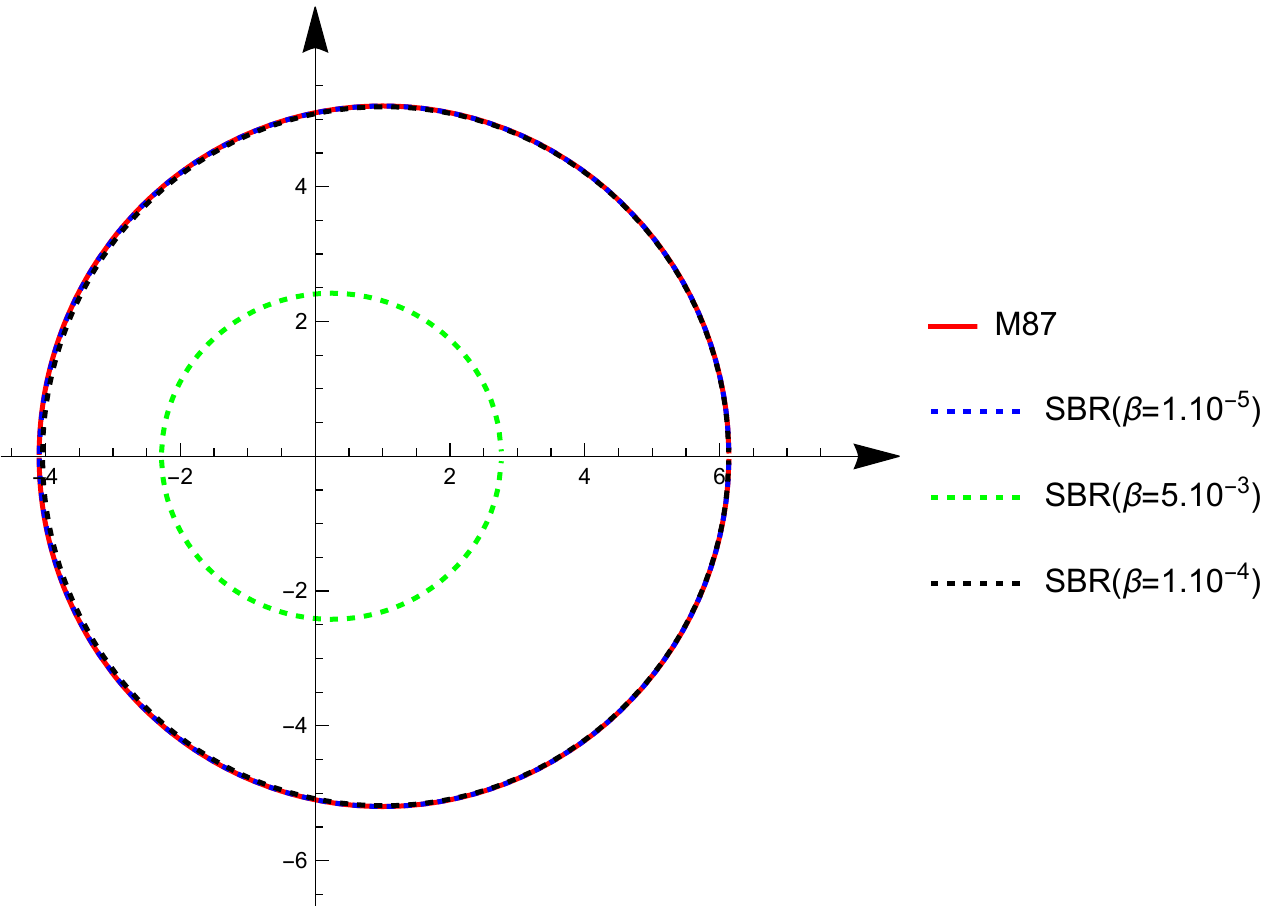}};
\node[] at (10,1){\small  \includegraphics[scale=0.60]{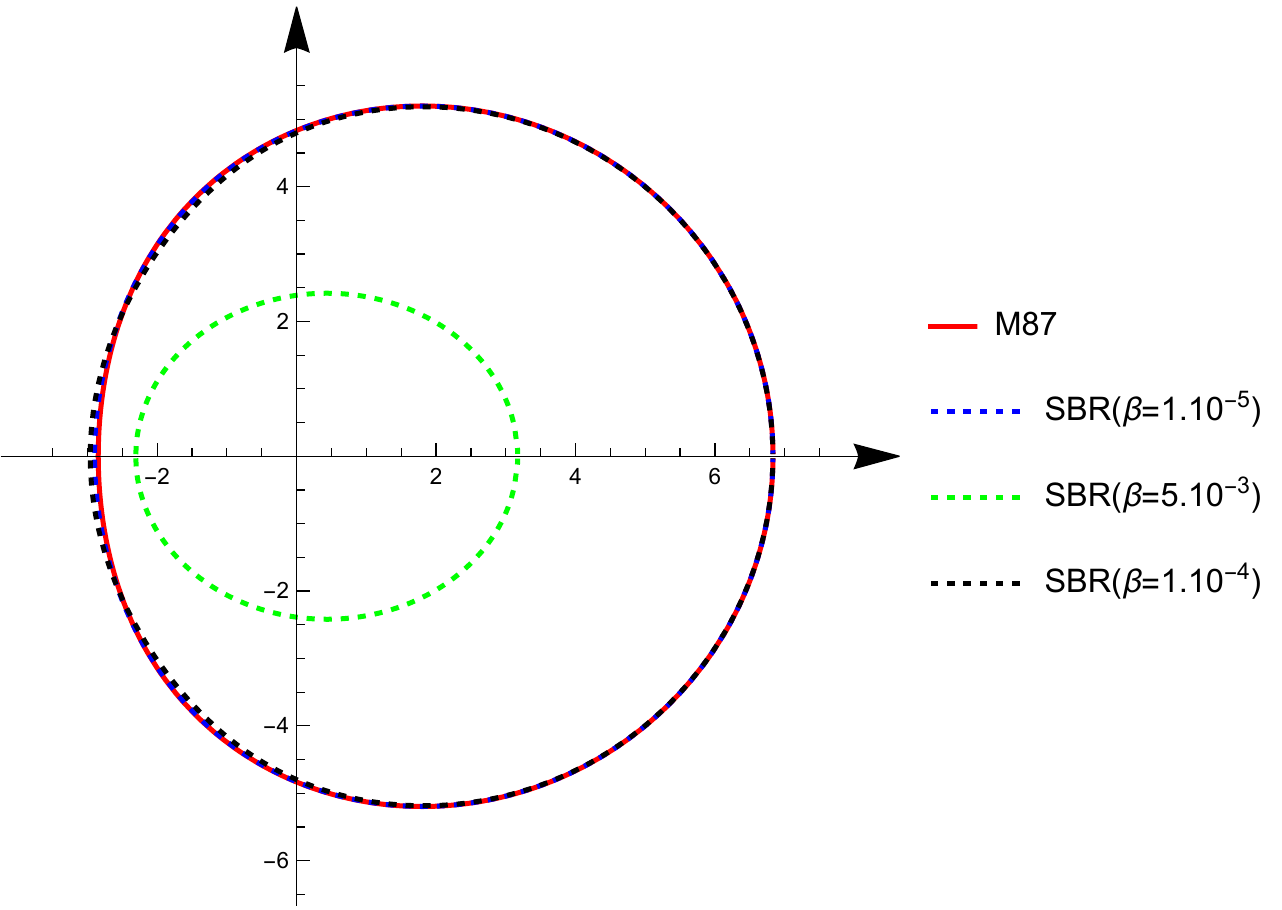}};
\node[color=black] at (-44,18.5) {$a = 0.5 $};
\node[color=black] at (0.5,18.5) {$a = 0.9$};
\node[color=black,rotate=90] at (-46.5,13) {\tiny{$Y_{SBR} $}};
\node[color=black,rotate=90] at (-55.5,0.7) {\tiny{$X_{SBR} $}};
\node[color=black,rotate=90] at (-2,13) {\tiny{$Y_{SBR} $}};
\node[color=black,rotate=90] at (-10.5,0.7) {\tiny{$X_{SBR} $}};
\end{tikzpicture}	
\caption{{\it \footnotesize SBR Black hole shadows for different values of $\beta$ and $a$ compared with $M87^\ast$. We take  the M87 black hole mass $M_{BH} = 6.5 \times 10^9M_\odot$ and $r_0 = 91.2kpc$.}}
\label{CN}
\end{center}
\end{figure*}
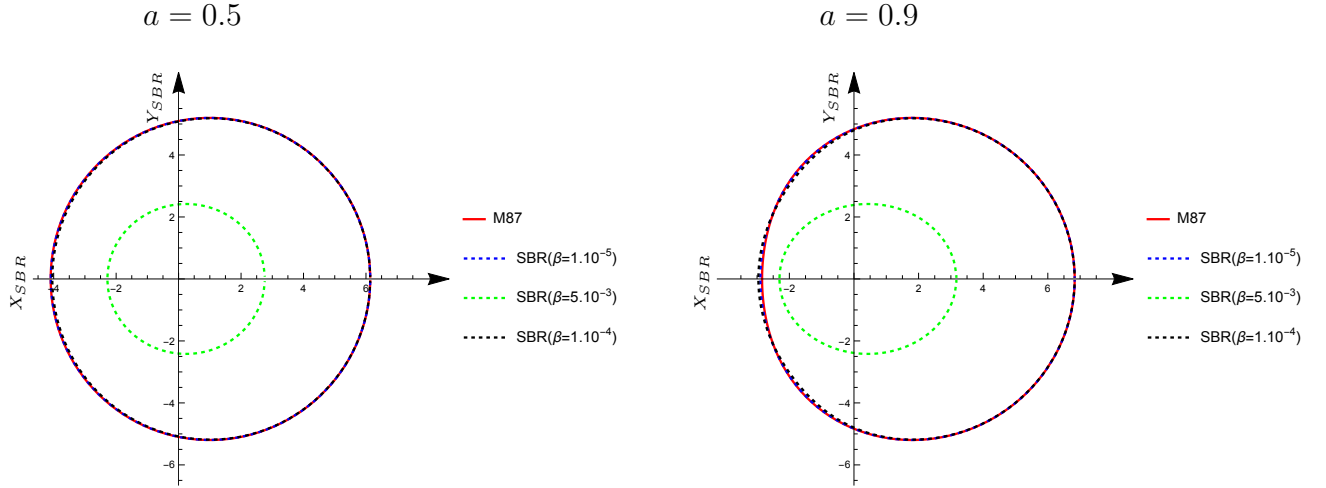

In this subsection, we would like to  constrain  the stringy  gravity parameter by means of EHT 
observational data. 
A close examination reveals that the  observational data   corresponding to   the shadow of the supermassive black hole $M87^*$, provided  by the   EHT   international  collaborations, could be exploited to   test and  probe the proposed gravity models.  It has been shown   that   such an  empiric exam   could    impose     constraints on the    black hole parameters  in question \cite{B12,V1,V2,V3,V4,V5,V6}.       In the unit of  the $M87^*$ mass,  we could   superpose  the $M87^\ast$  black hole shadow  behaviors    given by  a Kerr solution  and the SBR black holes. Fixing the rotating parameter, we can confront the present  gravity model   with  the  observational fundings.  In Fig.(\ref{CN}), we give two shadow configurations  associated with  two rotating parameter values.   
It has been  observed  that the rotating parameter  and  the $M87^\ast$  shadow data could   fix the  stringy gravity parameter  $\beta$ originated from  the M-theory compactification.   
A close comparison between the  shadow radius of   the SBR and  the $M87^\ast$ black holes conforms that     the SBR shadows could coincide perfectly with the $M87^\ast$ ones for  specific values of   the stringy gravity parameter.  The possible compatibilities impose the following  constraint  on $\beta$
\begin{equation}
\beta< 10^{-4}.
\end{equation}
Alternative studies  associated with  the   cosmological  building models in the SBR   gravity support such a constraint \cite{SBR1,SBR2,SBR3}.
For   $\beta>10^{-4}$,  however, we remark   distinctions between  both black hole shadows.   In what follows,  we consider such constraints to investigate   the deflection  angle of the  light rays near to the proposed  rotating black holes in the   SBR  gravity.

\section{ Light ray  behaviors near  to SBR  black holes}
In this section, we  investigate  the light ray  behavior near  to the  SBR black holes  by considering  the deflection angle  using the Gauss Bonnet theorem results \cite{BW18}.  Precisely, we would  like to  examine  the  impact of  the stringy   gravity parameter $\beta$  on the deflection angle of  the light rays  by  the SBR black holes.   In the equatorial plane, the deflection angle can be expressed
\begin{equation}
 \alpha=\Psi_{R}-\Psi_{S}+\phi_{SR}
 \label{a1}
 \end{equation}
where  $\Psi_{R}$ and $\Psi_{S}$    are the  angles between the light rays and the radial direction at the observer and  the source positions,  respectively.     It is denoted  that $\phi_{SR}$   is the longitude separation angle \cite{BW27}.   To obtain such an optical quantity, we could adopt the algorithm developed in \cite{BW23}. 
Taking the following line element 
\begin{equation}
ds^2=-A(r,\theta)dt^2+B(r,\theta)dr^2+C(r,\theta)d\theta^2+D(r,\theta)d\phi^2-2H(r,\theta)dt d\phi,
\end{equation}
we can compute the above angular  quantities  \cite{BW23}. 	The $\psi$ terms can be  extracted from the following equation
\begin{equation}
\label{G111}
\sin\Psi=\frac{H(r)+A(r)b}{\sqrt{A(r)D(r)+H^2(r)}}
\end{equation}
where $b$ is  identified  with  the impact  parameter  $\frac{L}{E}.$
Placing  the observer (R) and the source (S)  at a 
finite distance and making the change of variable $u=\frac{1}{r}$,  the separation angle  takes the form 
\begin{equation}
\phi_{RS} = \int^R_S d\phi= \int^{u_0}_{u_S}\frac{1}{\sqrt{F(u)}}du +\int^{u_0}_{u_R}\frac{1}{\sqrt{F(u)}}du
\end{equation} 
 $u_{S}$ and $u_{R}$ denote  the inverse of the source and the observer distance from the black hole, respectively. 
 $u_{0}$ is the inverse of the closest approach $r_{0}$.  The $F(u)$ function  is expressed as  follows
\begin{eqnarray}
F(u)=\left(\frac{1}{u^2} \frac{du}{d\phi}\right)^2. 
\end{eqnarray}
The computations lead to 
\begin{eqnarray}
F(u)= \frac{u^4 \left(A\left({u}\right) D\left({u}\right)+H\left({u}\right)^2\right) \left(b^2 \left(-A\left({u}\right)\right)-2 b H\left({u}\right)+D\left({u}\right)\right)}{B\left({u}\right) \left(b A\left({u}\right)+H\left({u}\right)\right)^2}.
\end{eqnarray}
Having given the essential on  the deflection angle terms, we move to compute  its  explicit expressions for both cases:  non-rotating and rotating solutions.  
\subsection{Deflection angle of non-rotating  solutions}
\begin{figure}
		\begin{center}
			\includegraphics[scale=0.9]{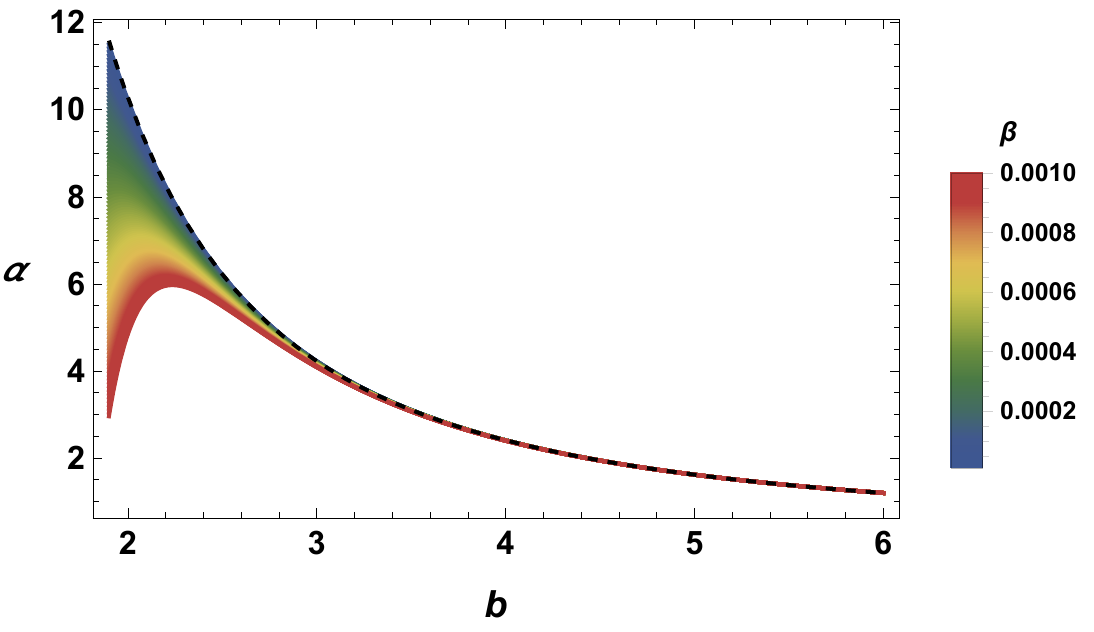} 
\caption{{\it \footnotesize Variation of  deflection angle of SBR black hole in terms of the impact parameter for large range of the string parameter. Dashed black  curve: variation of deflection angle of Schwarzschild  black hole.}}
\label{f111}
\end{center}
\end{figure}

To start the deflection angle computations,  we consider the non-rotating solutions of  the SBR black holes.   The calculations give 
\begin{eqnarray}
\phi_{RS}&=&\left(\pi-\arcsin \left(b u_R\right)-\arcsin \left(b u_S\right) \right)+\left(\frac{2-b^2 u_R^2}{\sqrt{1-b^2 u_R^2}}+\frac{2-b^2 u_S^2}{\sqrt{1-b^2 u_S^2}}\right)\frac{G M}{b}\notag\\&+&\left(\pi-\arcsin \left(b u_R\right)-\arcsin \left(b u_S\right) \right)\frac{15G^2 M^2}{4 b^2} \notag \\&+&\left(\frac{u_R \left(3 b^4 u_R^4-20 b^2 u_R^2+15\right)}{\left(1-b^2 u_R^2\right){}^{3/2}}+\frac{u_S \left(3 b^4 u_S^4-20 b^2 u_S^2+15\right)}{\left(1-b^2 u_S^2\right){}^{3/2}}\right)\frac{G^2 M^2}{4 b^2}\notag \\&+& \left(\frac{-5 b^8 u_R^8-40 b^6 u_R^6+240 b^4 u_R^4-320 b^2 u_R^2+128}{\left(1-b^2 u_R^2\right){}^{5/2}}\right.\\&+&\left.{}\frac{-5 b^8 u_S^8-40 b^6 u_S^6+240 b^4 u_S^4-320 b^2 u_S^2+128}{\left(1-b^2 u_S^2\right){}^{5/2}}\right)\frac{G^3 M^3}{6 b^3}\notag \\&+&\notag\left(\frac{7 b^{10} u_R^{10}+10 b^8 u_R^8+16 b^6 u_R^6+32 b^4 u_R^4+128 b^2 u_R^2-256}{\sqrt{1-b^2 u_R^2}}\right.\\&+&\left.{}\frac{7 b^{10} u_S^{10}+10 b^8 u_S^8+16 b^6 u_S^6+32 b^4 u_S^4+128 b^2 u_S^2-256}{\sqrt{1-b^2 u_S^2}}\right)\frac{6144 \pi ^3 \beta  G^6 M^3}{35 b^9}.\notag
\end{eqnarray}
For the $\Psi$ expressions,  we find 
\begin{eqnarray}
\Psi_{R}-\Psi_{S}&=&\left(\arcsin \left(b u_R\right)+\arcsin \left(b u_S\right)-\pi \right)-\left(\frac{u_R^2}{\sqrt{1-b^2 u_R^2}}+\frac{u_S^2}{\sqrt{1-b^2 u_S^2}}\right)b G M \notag \\&+&  \left(\frac{u_R^3 \left(2 b^2 u_R^2-1\right)}{\left(1-b^2 u_R^2\right){}^{3/2}}+\frac{u_S^3 \left(2 b^2 u_S^2-1\right)}{\left(1-b^2 u_S^2\right){}^{3/2}}\right)\frac{b G^2 M^2}{2} +\left(\frac{u_R^4 \left(-8 b^4 u_R^4+8 b^2 u_R^2-3\right)}{\left(1-b^2 u_R^2\right){}^{5/2}}\right.\\&+&\left.{}\frac{u_S^4 \left(-8 b^4 u_S^4+8 b^2 u_S^2-3\right)}{\left(1-b^2 u_S^2\right){}^{5/2}}\right)\frac{ b G^3 M^3}{6}+\left(\frac{u_R^{10}}{\sqrt{1-b^2 u_R^2}}+\frac{u_S^{10}}{\sqrt{1-b^2 u_S^2}}\right)\frac{55296\pi ^3 b \beta  G^6 M^3}{5}. \notag
\end{eqnarray}
Combining   the obtained  expressions, we  get 
\begin{eqnarray}
\alpha&=& \left(\sqrt{1-b^2 u_R^2}+\sqrt{1-b^2 u_S^2}\right)\frac{2 G M}{b}+\left(\pi-\arcsin \left(b u_R\right)-\arcsin \left(b u_S\right) \right)\frac{15 G^2 M^2}{4 b^2}\notag \\&+& \left(\frac{15 b u_R-7 b^3 u_R^3}{\sqrt{1-b^2 u_R^2}}+\frac{15 b u_S-7 b^3 u_S^3}{\sqrt{1-b^2 u_S^2}}\right)\frac{G^2 M^2}{4 b^2}\nonumber\\&+&\left(\frac{13 b^6 u_R^6+45 b^4 u_R^4-192 b^2 u_R^2+128}{\left(1-b^2 u_R^2\right){}^{3/2}}+\frac{13 b^6 u_S^6+45 b^4 u_S^4-192 b^2 u_S^2+128}{\left(1-b^2 u_S^2\right){}^{3/2}}\right)\frac{G^3 M^3}{6 b^3} \nonumber\\&+&\left(\frac{35 b^{10} u_R^{10}+5 b^8 u_R^8+8 b^6 u_R^6+16 b^4 u_R^4+64 b^2 u_R^2-128}{\sqrt{1-b^2 u_R^2}}\right.\\&+&\left.{}\frac{35 b^{10} u_S^{10}+5 b^8 u_S^8+8 b^6 u_S^6+16 b^4 u_S^4+64 b^2 u_S^2-128}{\sqrt{1-b^2 u_S^2}}\right)\frac{12288 \pi ^3 \beta  G^6 M^3}{35 b^9}.  \nonumber
\end{eqnarray}
Sending   $u_R$ and $ u_S$ to zero, we can  obtain an approximated expression of the deflection angle of  the light rays   near  to the non-rotating SBR black holes. Indeed, it is given by   
\begin{eqnarray}
\alpha &\simeq&\frac{4 G M}{b}+\frac{15 \pi  G^2 M^2}{4 b^2}+\frac{128 G^3 M^3}{3 b^3}-\frac{3145728 \pi ^3 \beta  G^6 M^3}{35 b^9}.
\end{eqnarray}
The deflection angle   depends on  $M$, $ b$  and  the stringy  gravity parameter $\beta$.  The previous shadow investigation has revealed that certain constraints on $\beta$ should be imposed.  Taking into account such conditions,  we  present  this  deflection angle of light rays    as a function of the impact parameter by varying the stringy gravity  parameter $\beta$.   Fig.(\ref{f111}) illustrates this variation for different values of $\beta$.
For small values of  $b$, we  remark   a  considerable   effect of $\beta$ on the deflection angle.  Precisely, it decreases by increasing $\beta$. For large values  of $b$,   we do not  observe  a relevant  distinction    due to coinciding curves of the deflection angle of the  light rays.  In this region of the moduli space, the deflection angle  decreases  slightly.  \subsection{Deflection angle of rotating  SBR black holes}
\begin{figure}[ht!]
		\begin{center}
		\centering
			\begin{tabbing}
			\centering
			\hspace{8.cm}\=\kill
			\includegraphics[scale=0.7]{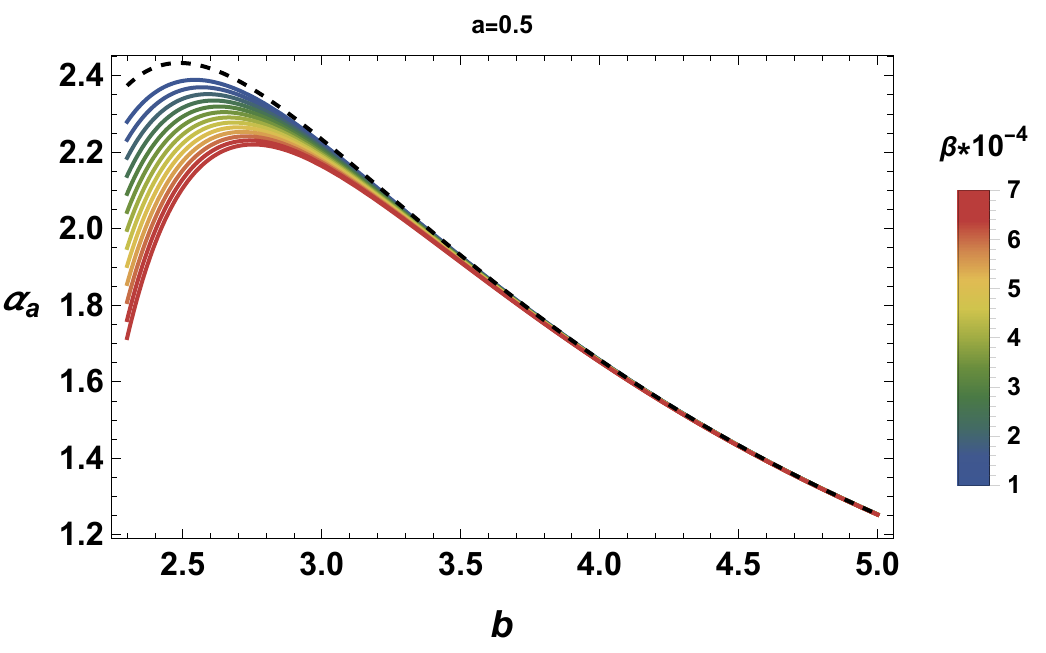} 
	\hspace{0.1cm}		\includegraphics[scale=0.7]{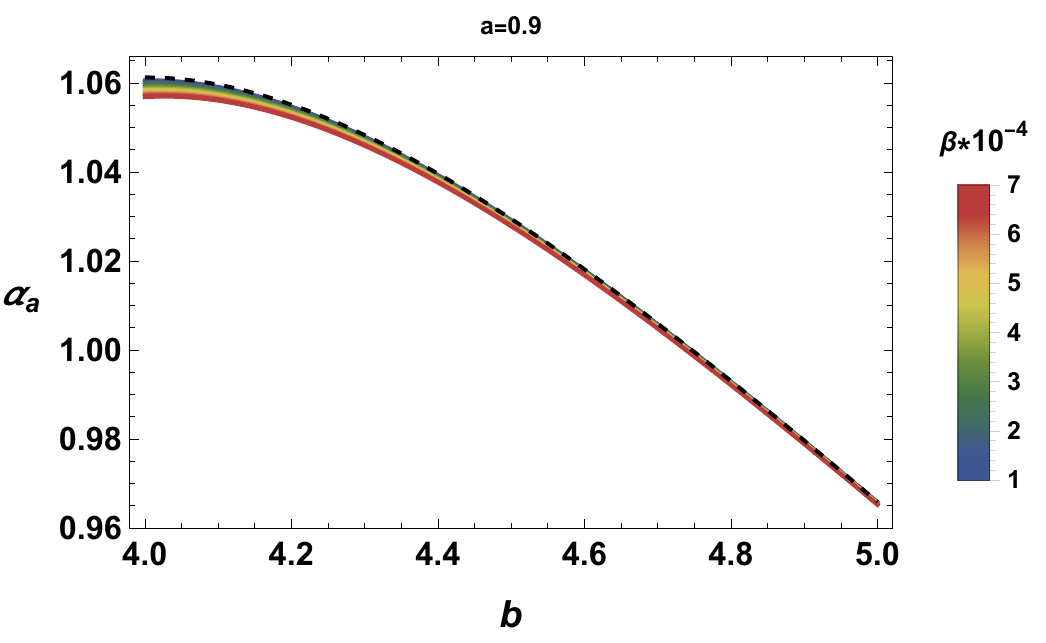}\\ 
	
	    \end{tabbing}
\caption{ {\it \footnotesize Deflection angle of the  light rays near  to the rotating SBR black holes as a function of the impact parameter   by   varying $\beta$ and  fixing $a$. }}
\label{def111}
\end{center}
\end{figure}

Here,   we investigate the impact of the stringy  gravity parameter $\beta$   on the deflection angle of  the light rays  by rotating  black holes in the   SBR gravity. Using similar calculations,  we can  obtain the following  separation angle for the slowly rotating black holes. Indeed,  we find  
\begin{eqnarray}
\phi_{RS_{a}}&=&\phi_{RS}-\left( \frac{1}{\sqrt{1-b^2 u_R^2}}+\frac{1}{\sqrt{1-b^2 u_S^2}}\right) \frac{2 G M a }{b^2}+\left(\arcsin \left(b u_R\right)+\arcsin \left(b u_S\right)-\pi \right)\frac{10  G^2 M^2 a}{b^3}
\notag\\ &+&\notag \left(\frac{b u_R \left(6 b^2 u_R^2-5\right)}{\left(1-b^2 u_R^2\right){}^{3/2}}+\frac{b u_S \left(6 b^2 u_S^2-5\right)}{\left(1-b^2 u_S^2\right){}^{3/2}}\right)\frac{2  G^2 M^2 a}{b^3}+\left(\frac{35 b^6 u_R^6-182 b^4 u_R^4+240 b^2 u_R^2-96}{\left(1-b^2 u_R^2\right){}^{5/2}}\right.\\&+&\left.{}\frac{35 b^6 u_S^6-182 b^4 u_S^4+240 b^2 u_S^2-96}{\left(1-b^2 u_S^2\right){}^{5/2}}\right)\frac{ G^3 M^3 a}{b^4}-\left(\frac{5 b^8 u_R^8+8 b^6 u_R^6+16 b^4 u_R^4+64 b^2 u_R^2-128}{\sqrt{1-b^2 u_R^2}}\notag \right.\\&+&\left.{}\frac{5 b^8 u_S^8+8 b^6 u_S^6+16 b^4 u_S^4+64 b^2 u_S^2-128}{\sqrt{1-b^2 u_S^2}}\right)\frac{110592 \pi ^3  \beta  G^6 M^3 a}{175 b^{10}}.
\end{eqnarray}
The remaining  deflection angle equation  can be expressed  as follows
\begin{eqnarray}
\Psi_{RS_{a}}&=&\Psi_{RS}+\left(\frac{u_R^2}{\sqrt{1-b^2 u_R^2}}+\frac{u_S^2}{\sqrt{1-b^2 u_S^2}}\right)2  G M a-\left(\frac{u_R^3 \left(2 b^2 u_R^2-1\right)}{\left(1-b^2 u_R^2\right){}^{3/2}}+\frac{u_S^3 \left(2 b^2 u_S^2-1\right)}{\left(1-b^2 u_S^2\right){}^{3/2}}\right)2  G^2 M^2 a\notag \\&+&\left(\frac{u_R^4 \left(8 b^4 u_R^4-8 b^2 u_R^2+3\right)}{\left(1-b^2 u_R^2\right){}^{5/2}}+\frac{u_S^4 \left(8 b^4 u_S^4-8 b^2 u_S^2+3\right)}{\left(1-b^2 u_S^2\right){}^{5/2}}\right) G^3 M^3 a\\&-&\left(\frac{u_R^{10}}{\sqrt{1-b^2 u_R^2}}+\frac{u_S^{10}}{\sqrt{1-b^2 u_S^2}}\right)\frac{110592}{5} \pi ^3  \beta  G^6 M^3 a \notag
\end{eqnarray}
where     the notation  $\Psi_{RS_{a}}= \Psi_{R_{a}}-\Psi_{S_{a}}$ has  been used. 
Using the above equations,  the expression of the deflection angle  of light rays of    the rotating SBR  black holes is found to be 
\begin{eqnarray}
\alpha_{RS_{a}}&=&\alpha_{RS}-\frac{2 a G M}{b^2} \left(\sqrt{1-b^2 u_R^2}+\sqrt{1-b^2 u_S^2}\right)+\frac{2 a G^2 M^2}{b^3} \left(\frac{2 b^3 u_R^3-5 b u_R}{\sqrt{1-b^2 u_R^2}}+\frac{2 b^3 u_S^3-5 b u_S}{\sqrt{1-b^2 u_S^2}}\right)\notag\\&+&\frac{10 a G^2 M^2}{b^3} \left(\arcsin\left(b u_R\right)+\arcsin\left(b u_S\right)-\pi \right)-\frac{a G^3 M^3}{b^4} \left(\frac{8 b^6 u_R^6+35 b^4 u_R^4-144 b^2 u_R^2+96}{\left(1-b^2 u_R^2\right){}^{3/2}}\notag \right.\\&+&\left.{}\frac{8 b^6 u_S^6+35 b^4 u_S^4-144 b^2 u_S^2+96}{\left(1-b^2 u_S^2\right){}^{3/2}}\right)\notag\\&+&\frac{110592 \pi ^3 a \beta  G^6 M^3}{175 b^{10}} \left(\sqrt{1-b^2 u_R^2} \left(35 b^8 u_R^8+40 b^6 u_R^6+48 b^4 u_R^4+64 b^2 u_R^2+128\right)\notag \right.\\&+&\left.{}\sqrt{1-b^2 u_S^2} \left(35 b^8 u_S^8+40 b^6 u_S^6+48 b^4 u_S^4+64 b^2 u_S^2+128\right)\right).
\end{eqnarray}
Taking  appropriate limits,  we  can  get    such the    expression of the deflection angle of light rays   in  the SBR  gravity.  Indeed,   it  takes the form   
\begin{eqnarray}
\alpha_{{a}}&\simeq&\frac{4 G M}{b}+\frac{15 \pi  G^2 M^2}{4 b^2}+\frac{128 G^3 M^3}{3 b^3}-\frac{3145728 \pi ^3 \beta  G^6 M^3}{35 b^9}-\frac{4 a G M}{b^2}-\frac{10 \pi  a G^2 M^2}{b^3}-\frac{192 a G^3 M^3}{b^4}\notag\\&+&\frac{28311552 \pi ^3 a \beta  G^6 M^3}{175 b^{10}}.
\end{eqnarray}
Considering  $a=0$, we  obtain  the deflection angle of  the non-rotating solutions obtained in the previous  section.   For  $\beta=0$, we  recover the case of  the Kerr solution.  By fixing the rotating parameter,  we  illustrate in  Fig.(\ref{def111})  the  deflection angle   as a function of   $b$  by varying  the stringy  gravity parameter $\beta$ in the range obtained by the above computations   associated with EHT  requirements. 
\begin{figure}[ht!]
		\begin{center}
		\centering
			\begin{tabbing}
			\centering
	\hspace{8.cm}\=\kill
			\includegraphics[scale=0.7]{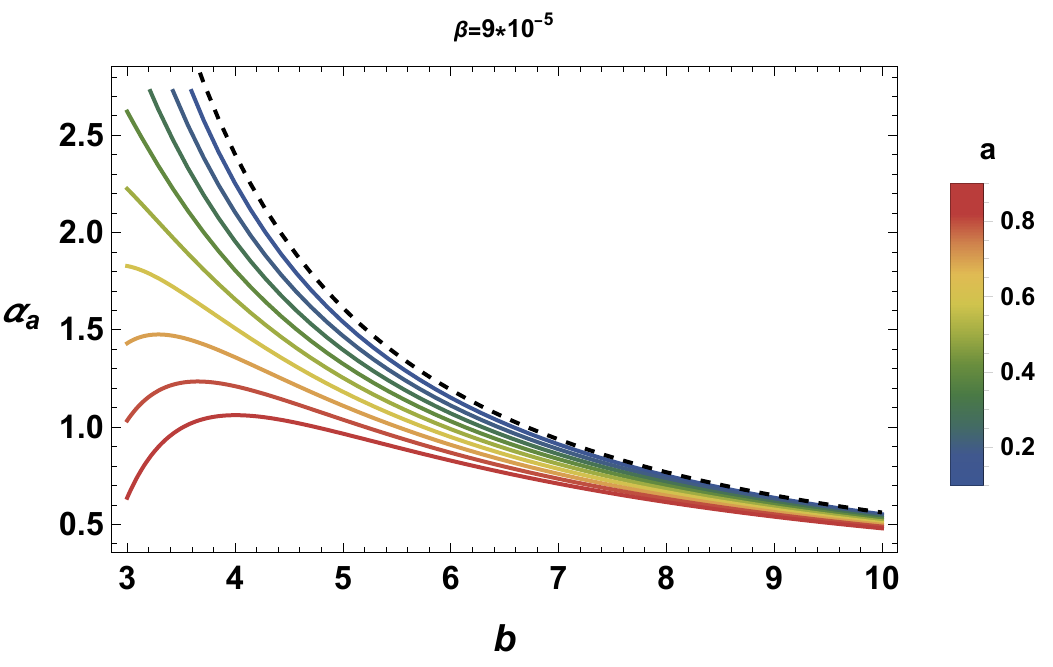} 
	\hspace{0.1cm}		\includegraphics[scale=0.7]{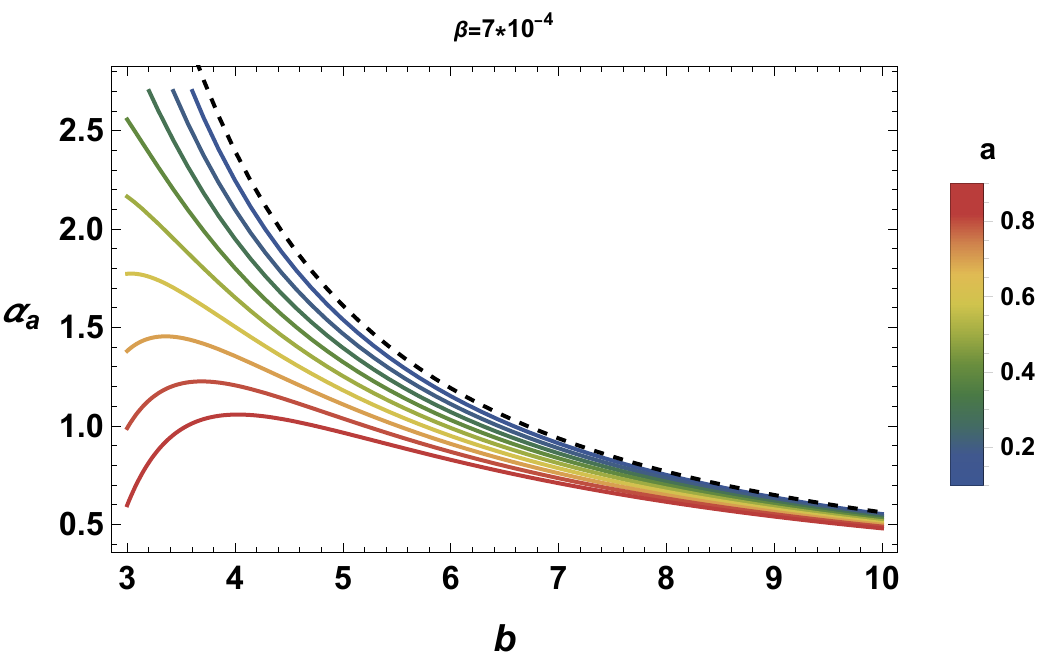}
	    \end{tabbing}
\caption{ {\it \footnotesize Deflection angle of light rays near rotating SBR black holes as a function of the impact parameter   by   varying $a$ and fixing $\beta$. }}
\label{da4}
\end{center}
\end{figure}

For small values of the impact parameter and  the rotating parameter,  we can observe    a relevant  effect of the stringy gravity  parameter which  decreases the deflection angle. It follows from  the figure that the deflection angle decreases with the rotating parameter.  For large values of the rotating parameter the $\beta$-curves are not  distinguishable, with respect to the Kerr ones. This could show that, for large rotating parameter values, the effect of $\beta$ becomes irrelevant. In  Fig.(\ref{da4}), we present  the variation of the deflection angle in terms of the impact parameter by varying the rotating parameter and fixing the stringy gravity parameter. Contrary to  the shadow  behavior,  the  two  fixed values of $\beta$  provide similar behaviors.  Moreover, the decreasing behavior of  the deflection angle depends on  both  the rotating parameter and  the stringy gravity one.  The associated critical impact parameter is shifted by increasing such parameters.

\section{Conclusion}

Motivated by M-theory compactifications, we  have studied   optical properties of  the black holes in  the    SBR   gravity.
  Precisely, we  have  discussed   the shadows and the deflection angle of  the light rays by non-rotating and rotating  black holes  in four dimensions.   In the first part of this work, we   have  investigated the shadows of the Schwarzschild-type solutions. As expected, we have   obtained    perfect circular shadows  where the size decreases with  the  stringy  gravity parameter  $\beta$.  The latter has been constrained by  $\beta<8.10^{-2}$  derived from  the shadow existence.  Indeed,  the radius of the horizon is small than  the shadow. For $\beta>8.10^{-2}$, however,  the shadow does not exist.  This  constraint  on $\beta$ matches  perfectly with certain findings associated with  the cosmological building  models in the SRB gravity\cite{SBR2}.  In such a range of the stringy gravity   parameter,  we have combined the  Newman-Janis algorithm and  the Hamilton-Jacobi mechanism to investigate the    shadow  behaviors of  the rotating solutions in terms of  one-dimensional real closed  curves.    More precisely,   we    have provided   various   sizes and shapes depending  on  the rotating and the stringy gravity parameters $a$  and $\beta$, resepctively.    To  examine the shadow geometric deformations,   we have  exploited the astronomical observables and   the energy emission rate.  As results,   we  have   shown    that the rotating parameter $a$  and   the  stringy gravity parameter $\beta$  could   control   the  shadow aspect.  For specific values of  $a$ and $\beta$, we  have observed    a pair of cusps  in the shadow geometry of  the SBR black holes.  It has been remarked that such a   pair of cusps  has appeared   in  the Kerr black hole solution in  plasma distribution backgrounds \cite{Z1,Z2}.    We  anticipate that  the  plasma and the  stringy gravity parameter could have a  quite similar effect on the  shadow shapes.  As mentioned before, this could deserve more understandings.   Using  EHT observational  data, we  have provided    constraints on    $\beta$   which could play a relevant role in M-theory compactifications being in  a good agreement with  results obtained  from  recent cosmological  investigations.  Taking into account these constraints, we have elaborated  a concise study  on the  light ray  behaviors  near to    the   SBR   black holes  in   the    SBR   gravity by computing the  deflection angle in terms of  $a$ and $\beta$. In particular,  we have presented    a  graphical  analysis  of such a deflection  angle of the  light rays.  We have observed that, for large values of the rotating parameter,  the stringy  gravity parameter does not affect the deflection angle of the light rays near to the  SBR black holes. \\
   This   work comes up with  open questions.   A natural question concerns the link with M-theory compactifications on non-trivial geometries including the G2-manifolds. It could be possible to use the  black hole physics  to control    the gravity parameters from     observational  data associated  either with $M87^*$ or  Sgr A$^*$ \cite{V7,V8}.  In this way,  the rotating black holes   can    play a primordial role in testing gravity  models   via its moduli space.  Moreover, the  SBR   gravity has received a special  interest in connections with inflation models \cite{inflation}. It should be interesting to implement  the coupling inflation scenario  in  such a gravity.  It  could be possible to  address  all these  issues  in future works.

\section*{Acknowledgements}
The authors  would like to thank  N. Askour,   H. El Moumni, S-E. Ennadifi,  M. Lamaaoune  M. Oualaid, and  Y. Sekhmani   for  discussions and  recent collaborations on related topics.


\begin{thebibliography}{100}
 \bibitem{3}
 B. Abbott and al., \textit{Observation of Gravitational Waves from a Binary Black Hole Merger},
Phys.\ Rev.\ Lett.\ {\bf 116} (6) (2016) 061102, {\tt arXiv:1602.03837}.
\bibitem{4}
K. Akiyama and al.,
\textit{First M87 Event Horizon Telescope Results. IV. Imaging the Central
  Supermassive Black Hole},
 Astrophys. J. {\bf L4} (1) (2019) 875, {\tt arXiv:1906.11241}.

\bibitem{5}
K. Akiyama and al.,
\textit{First M87 Event Horizon Telescope Results. V. Imaging the Central
  Supermassive Black Hole},
 Astrophys. J. {\bf L5} (1) (2019) 875.
\bibitem{6}
K. Akiyama and al.,
\textit{First M87 Event Horizon Telescope Results. VI. Imaging the Central
  Supermassive Black Hole},
 Astrophys. J. {\bf L6} (1) (2019) 875.
  \bibitem{R5}
  S.W. Hawking, H.S. Reall,{\it Charged and rotating AdS black holes and their CFT duals}, Phys.Rev.D  \textbf{61} (2000) 024014, {\tt arXiv:hep-th/9908109}.
   \bibitem{R6}
   A. Chamblin, R. Emparan, C. V. Johnson, R. C. Myers,{\it Charged AdS Black Holes and Catastrophic Holography},Phys.Rev.D \textbf{60} (1999) 064018,  {\tt arXiv:hep-th/9902170}.
 \bibitem{19}
 A. Rajagopal,  D. Kubiznak,  R. B. Mann,  \textit{Van der Waals black hole}, Phys. Lett. B\textbf{737} (2014) 277, {\tt arXiv:1408.1105}.
 \bibitem{Ref10}
D. Kubiznak, R. B. Mann, Mae Teo, Black hole chemistry: thermodynamics with
Lambda, Class. Quant.  Grav. {\bf 34} (2017) 063001,  {\tt arXiv:1608.06147}.
 \bibitem{Ref7}
S. W. Hawking, D. N. Page, Thermodynamics of black holes in anti-de Sitter space,
Communications in Mathematical Physics {\bf 87}(4) (1983) 577.
\bibitem{Ref5}  
G.W. Gibbons, M.J. Perry, C.N. Pope,  The First Law of Thermodynamics for Kerr-Anti-de Sitter Black Holes,   	Class. Quant. Grav. {\bf 22} (2005)1503.
\bibitem{18}
A. Belhaj, M. Chabab, H. El Moumni and M. B. Sedra,  \textit{On thermodynamics of AdS black
holes in arbitrary dimensions}, Chin. Phys. Lett. \textbf{29} (2012) 100401.
\bibitem{181}
 F.  Barzi, H. El Moumni,  {\it 
On Rényi universality formula of charged flat black holes from Hawking-Page phase transition},  Phys. Lett. B,
 \textbf{ 833}(2022)137378.
\bibitem{182}
A.  Belhaj, A.  El Balali, W.  El Hadri, and E.  Torrente-Lujan, {\it On universal
constants of AdS black holes from Hawking-Page phase transition}, Phys. Lett. B.
 \textbf{ 811}(2020)135871.
\bibitem{Ref6}
R. Banerjee, S. Ghosh and D. Roychowdhury, {\it  New type of phase transition in Reissner
Nordstrm AdS black hole and its thermodynamic geometry,} Phys. Lett.B  {\bf 696} (2011) 156.

 \bibitem{17} Y. Liu, D. C. Zou,  B. Wang, \textit{Signature of the Van der Waals like small-large charged AdS black hole phase transition in quasi normal modes}, JHEP \textbf{09} (2014) 179, {\tt arXiv:1405.2644}.
 

\bibitem{180}
A. Belhaj, M. Chabab, H. El Moumni, K. Masmar, M. B. Sedra,  \textit{On thermodynamics of AdS
black holes in M-theory}, Eur. Phys. J. C \textbf{76}(2) (2016) 73.

 



 


\bibitem{9}
 A. Belhaj, H. Belmahi,  M. Benali, W. El Hadri, H. El Moumni, E. Torrente-Lujan,  \textit{Shadows of 5D Black Holes from string theory},  Phys. Lett.
B  \textbf{812} (2021) 136025, {\tt arXiv:2008.13478}.
\bibitem{10}
A.~Belhaj, H.~Belmahi,  M.~Benali, {\it Superentropic AdS black hole shadows},
Phys. Lett. B \textbf{821} (2021) 136619, {\tt arXiv:2110.06771}.

 \bibitem{d1M}
A. Belhaj, M. Benali, A. El Balali, H. El Moumni and S-E. Ennadifi,
\textit{Deflection angle and shadow behaviors of quintessential black holes in arbitrary dimensions}, Class. Quantum Grav. \textbf{37} (2020) 215004, {\tt arXiv:2006.01078}.
\bibitem{d2}
W. Javed, J. Abbas, and  A. Övgün,  \textit{Deflection angle of photon from magnetized black hole and effect
of nonlinear electrodynamics}, Eur. Phys. J. C  \textbf{79}  (2019) 694, {\tt arXiv:1908.09632}.

\bibitem{d3H} 
A. Belhaj, H. Belmahi, M. Benali, {\it Deflection Light Behaviors by AdS Black Holes},  Gen.Rel.Grav. \textbf{79}  54 (2022) 4,  {\tt arXiv:2112.06215}.
\bibitem{d4H}
A. Belhaj, H. Belmahi, M. Benali, H. El Moumni, {\it Light Deflection by Rotating Regular Black Holes with a Cosmological Constant}, {\tt arXiv:2204.10150}.
\bibitem{d51}
W. Javed, J. Abbas, and  A. Övgün,  \textit{Deflection angle of photon from magnetized black hole and effect
of nonlinear electrodynamics}, Eur. Phys. J. C \textbf{79}  (2019) 694, {\tt arXiv:1908.09632}.

\bibitem{Belhaj2}
A.~Belhaj, H.~Belmahi, M.~Benali and H.~El Moumni, {\it Light deflection angle by superentropic black holes}, Int. J. Mod. Phys. D \textbf{31} (2022) 2250054, {\tt arXiv:2203.11143}.
\bibitem{Belhaj3}
A.~Belhaj, H.~Belmahi, M.~Benali and A.~Segui, {\it Thermodynamics of AdS black holes from deflection angle formalism}, Phys. Lett. B \textbf{817} (2021) 136313.
\bibitem{Carlo1}
C.~A.~R.~Herdeiro, A.~M.~Pombo, E.~Radu, P.~V.~P.~Cunha and N.~Sanchis-Gual, {\it The imitation game: Proca stars that can mimic the Schwarzschild shadow}, JCAP \textbf{04} (2021) 051, {\tt arXiv:2102.01703}. 

\bibitem{Carlo2}
C.~A.~R.~Herdeiro, A.~M.~Pombo, E.~Radu, P.~V.~P.~Cunha and N.~Sanchis-Gual, {\it The imitation game: Proca stars that can mimic the Schwarzschild shadow}, JCAP \textbf{04} (2021) 051, {\tt arXiv:2102.01703}. 
\bibitem{K}
S. W. Wei, Y. C. Zou, Y. X. Liu, R. B. Mann, \textit{Curvature radius and Kerr black hole
shadow}, JCAP \textbf{08} (2019) 030, {\tt arXiv:1904.07710}.

\bibitem{BC}
J.~R.~Farah, D.~W.~Pesce, M.~D.~Johnson, L.~L.~Blackburn, {\it On the approximation of the black hole shadow with a simple polar curve}, Astrophys. J. \textbf{900} (2020)  77,  {\tt arXiv:2007.06732}.
\bibitem{Xa}
S.~V.~M.~C.~B.~Xavier, P.~V.~P.~Cunha, L.~C.~B.~Crispino, C.~A.~R.~Herdeiro,
\textit{Shadows of charged rotating black holes: Kerr\textendash{}Newman versus Kerr\textendash{}Sen},
Int. J. Mod. Phys. D \textbf{29}  (2020)  2041005, 
{\tt arXiv:2003.14349}.
\bibitem{J}
S. U. Khan, J. Ren, \textit{Shadow cast by a rotating charged black hole in quintessential dark energy}, Phys. Dark Univ. \textbf{30} (2020) 100644, {\tt arXiv:2006.11289}.
\bibitem{RC}
X. Hou, Z. Xu, J. Wang, \textit{Rotating black hole shadow in perfect
fluid dark matter}, JCAP \textbf{12} (2018) 040.
\bibitem{Belhaj4}
A.~Belhaj, M.~Benali and Y.~Hassouni, {\it Superentropic black hole shadows in arbitrary dimensions}, Eur. Phys. J. C \textbf{82} (2022) 619, {\tt arXiv:2203.06774}.
\bibitem{Belhaj11}
A.~Belhaj, H.~Belmahi and M.~Benali, {\it Superentropic AdS black hole shadows}, Phys. Lett. B \textbf{821} (2021) 136619, {\tt arXiv:2110.06771}.

\bibitem{B12}
A.~Belhaj, M.~Benali, A.~El Balali, W.~El Hadri, H.~El Moumni, E.~Torrente-Lujan,
{\it Black hole shadows in M-theory scenarios},
Int. J. Mod. Phys. D \textbf{30} (2021) 2150026, {\tt arXiv:2008.09908}.
\bibitem{ma}
A. Belhaj, A. El Balali, W. El Hadri, Y. Hassouni, E.~Torrente-Lujan, {\it Phase transition and shadow behaviors of quintessential black holes in M-theory/superstring inspired models}, Int. J. Mod. Phys. A  \textbf{36} (2021) 2150057,   {\tt arXiv:2004.10647}.


\bibitem{hajar}
N. Askour, A. Belhaj, H. Belmahi, M. Benali, H. El Moumni, Y. Sekhmani, {\it Light Behaviors around Black Holes in M-theory},   {\tt  arXiv:2301.08321}.

\bibitem{23Y}

J. T. Wheeler, \textit{Symmetric solutions to the Gauss-Bonnet extended Einstein equations}, Nucl. Phys. B\textbf{268}  (1986) 737.
\bibitem{230Y}  S. G. Ghosh, R. Kumar, \textit{Generating black holes in 4D Einstein-Gauss-Bonnet gravity}, Class.Quant.Grav. \textbf{37} (2020) 245008,  {\tt arXiv:2003.12291}.
\bibitem{24}  S. G. Ghosh,  S. D. Maharaj, \textit{Radiating black holes in the novel 4D Einstein-Gauss-Bonnet gravity}, Phys. Dark Univ. \textbf{30} (2020) 100687, {\tt arXiv:2003.09841}. 
 
\bibitem{282Y}
S. G. Ghosh, D. V. Singh, R. Kumar, S. D. Maharaj,   \textit{Phase transition of AdS black holes in 4D EGB gravity coupled to nonlinear electrodynamics},   Annals of Physics
{\bf 424}(2021)168347, {\tt  arXiv:2006.00594}.
 \bibitem{283Y} D. V. Singh, B. K. Singh, S. Upadhyay,   \textit{4D AdS Einstein–Gauss–Bonnet black hole with Yang–Mills field and its thermodynamics}, Annals of Physics {\bf 434}(2021) 168642.
 \bibitem{2855} A. Belhaj,  Y. Sekhmani,  {\it 
 Optical and thermodynamic behaviors of Ayón–Beato–García black holes for 4D Einstein Gauss–Bonnet gravity},  Annals of Physic {\bf 441}(2022)168863.
 
 \bibitem{285} A. Belhaj,  Y. Sekhmani,  {\it Thermodynamics of Ayon-Beato–Garcia–AdS black holes in 4D Einstein–Gauss–Bonnet gravity},   Eur. Phys. J. Plus {\bf 137}(2022)278.
  \bibitem{Sunny}
 S. Vagnozzi, R.  Roy, Yu-Dai Tsai, L.  Visinelli, M. Afrin, A. Allahyari, P. Bambhaniya, D.  Dey, S.  G. Ghosh, P.  S. Joshi, K.  Jusufi, M.  Khodadi, R. K.  Walia, A.  Övgün, C.  Bambi, {\it Horizon-scale tests of gravity theories and fundamental physics from the Event Horizon Telescope image of Sagittarius A$^*$},  {\tt   arXiv:2205.07787}.


 

 
\bibitem{SBR1}
S.  V. Ketov, {\it Starobinsky-Bel-Robinson gravity}, Universe \textbf{8} (2022) 351, {\tt  arXiv:2205.13172}. 
\bibitem{SBR2}
S. V. Ketov, E. O. Pozdeeva, S. Yu. Vernov, {\it On the superstring-inspired quantum correction to the Starobinsky model of inflation
}, JCAP \textbf{12} (2022) 032,  {\tt  arXiv:2211.01546}. 
\bibitem{SBR3}
R. C. Delgado, S. V. Ketov, {\it Schwarzschild-type black holes in Starobinsky-Bel-Robinson gravity}, Phys. Lett. B  \textbf{838}   (2023) 137690, {\tt arXiv:2209.01574}.
\bibitem{Z1}
Z.~Zhang, H.~Yan, M.~Guo and B.~Chen, {\it Shadows of Kerr black holes with a Gaussian-distributed plasma in the polar direction}, Phys. Rev. D \textbf{107} (2023) 024027, {\tt arXiv:2206.04430}.

\bibitem{Z2}
Y.~Huang, Y.~P.~Dong and D.~J.~Liu, {\it Revisiting the shadow of a black hole in the presence of a plasma}, Int. J. Mod. Phys. D \textbf{27} (2018) 1850114, {\tt arXiv:1807.06268}.

\bibitem{MT}
E. Witten,  {\it Solutions of four-dimensional field theories via M-theory}, Nucl. Phys. B {\bf 500} (1997) 42.
\bibitem{JNA}
S. P. Drake P. Szekeres, {\it An explanation of the Newman-Janis Algorithm}, Gen.Rel.Grav. \textbf{32} (2000) 445, {\tt arXiv:gr-qc/9807001}.
\bibitem{JNABH}
Harold Erbin,{\it Janis-Newman algorithm: generating rotating and NUT charged black holes},Universe \textbf{3} (2017)  19, {\tt arXiv:1701.00037}.

\bibitem{em}
 S. W. Wei and Y. X. Liu, {\it Observing the shadow of Einstein-Maxwell-Dilaton-Axion
black hole}, JCAP \textit{11} (2013) 063, {\tt arXiv:1311.4251}.

\bibitem{Belhaj5}
A.~Belhaj, M.~Benali, H.~El Moumni, M.~A.~Essebani, M.~B.~Sedra and Y.~Sekhmani, {\it Thermodynamic and optical behaviors of quintessential Hayward-AdS black holes}, Int. J. Geom. Meth. Mod. Phys. \textbf{19} (2022)  2250096,  {\tt  arXiv:2202.06290}.
\bibitem{Belhaj1}
A.~Belhaj, M.~Benali, A.~E.~Balali, W.~E.~Hadri and H.~El Moumni, {\it Cosmological constant effect on charged and rotating black hole shadows}, Int. J. Geom. Meth. Mod. Phys. \textbf{18} (2021) 2150188, {\tt arXiv:2007.09058}.
\bibitem{Atamurotov1} 
F.~Atamurotov and B.~Ahmedov, {\it Optical properties of black hole in the presence of plasma: shadow}, Phys. Rev. D \textbf{92} (2015) 084005, {\tt arXiv:1507.08131}.
\bibitem{Babar1}
G.~Z.~Babar, A.~Z.~Babar and F.~Atamurotov, {\it Optical properties of Kerr\textendash{}Newman spacetime in the presence of plasma}, Eur. Phys. J. C \textbf{80} (2020) 761, {\tt arXiv:2008.05845}.

\bibitem{V1} C. Bambi, K.  Freese, S.  Vagnozzi, L.  Visinelli,  {\it Testing the rotational nature of the supermassive object M87* from the circularity and size of its first image},  	Phys. Rev. D  \textbf{100} (2019) 044057,   	{\tt arXiv:1904.12983}.
\bibitem{V2} S.  Vagnozzi, L.  Visinelli,  {\it Hunting for extra dimensions in the shadow of M87$^*$},  Phys. Rev. D  \textbf{100} (2019) 024020,  	{\tt arXiv:1905.12421}.
\bibitem{V3} A.  Allahyari, M.  Khodadi, S. Vagnozzi, D. F. Mota, {\it  Magnetically charged black holes from non-linear electrodynamics and the Event Horizon Telescope},  	JCAP \textbf{2002} (2020) 003,   {\tt 	arXiv:1912.08231}
\bibitem{V4} M.  Khodadi, A. Allahyari, S.  Vagnozzi, D.  F. Mota,    {\it Black holes with scalar hair in light of the Event Horizon Telescope},   	JCAP   \textbf{2009} (2020) 026,   	arXiv:2005.05992 
\bibitem{V5} R. Roy, S.  Vagnozzi, L.  Visinelli,  {\it Superradiance evolution of black hole shadows revisited}, Phys. Rev. D \textbf{05} (2022) 083002, 	{\tt arXiv:2112.06932}. 
\bibitem{V6} S.  Vagnozzi, C.  Bambi, L.  Visinelli,  {\it  Concerns regarding the use of black hole shadows as standard rulers},   	Class. Quant. Grav. \textbf{37} (2020) 087001,  	{\tt arXiv:2001.02986}.



\bibitem{BW18}
T. Ono, A. Ishihara,  H. Asada, \textit{Gravitomagnetic bending angle of light with finite-distance corrections in stationary axisymmetric spacetimes}, Phys. Rev. D \textbf{96} (2017) 104037, {\tt  arXiv:1704.05615}. 
\bibitem{BW27}
R.~C.~Pantig and E.~T.~Rodulfo, {\it Weak deflection angle of a dirty black hole}, Chin. J. Phys. \textbf{66} (2020) 691, {\tt arXiv:2003.00764}.

\bibitem{BW23}
A.~Ishihara, Y.~Suzuki, T.~Ono, T.~Kitamura, H.~Asada, {\it Gravitational bending angle of light for finite distance and the Gauss-Bonnet theorem}, Phys. Rev. D \textbf{94} (2016) 084015, {\tt arXiv:1604.08308}.
\bibitem{V7}  Y.  Chen, R.  Roy, S.  Vagnozzi, L.  Visinelli,  {\it Superradiant evolution of the shadow and photon ring of Sgr A$^*$},  	Phys. Rev. D \textbf{106} (2022) 043021,   	{\tt arXiv:2205.06238}. 
\bibitem{V8} M. Afrin, S.  Vagnozzi, S. G. Ghosh,   {\it Tests of Loop Quantum Gravity from the Event Horizon Telescope Results of Sgr A$^*$},  	Astrophys. J. \textbf{944} (2023) 149,  {\tt 	arXiv:2209.12584}.
\bibitem{inflation}
 T.  Q. Do, D.  H. Nguyen, T.  M. Pham,  {\it Stability investigations of isotropic and anisotropic exponential inflation in the Starobinsky-Bel-Robinson gravity}, {\tt  arXiv:2303.17283}.  

\end{thebibliography}
\end{document}